\documentclass[]{article}
\usepackage{aas2pp4,epsf,lscape,latexsym}
\def\lastrev{Comments from TV,KG,ML,DA,DB,AB 21 March 2000}

\def\that{{\hat t}}
\def\umin{u_{min}}
\def\Amax{A_{max}}
\def\tmax{t_{max}}
\def\rthat{{\hat t'}}
\def\rumin{u_{min}'}
\def\rAmax{A_{max}'}
\def\rtmax{t_{max}'}

\def\ee#1{\times 10^{#1}}
\def\Ethat{{\cal E}(\that)}

\def\Et0{{\cal E}(\t0)}

\def\sodophot{{SoDoPhot}}
\def\dophot{{DoPhot}}
\def\daophot{{DaoPhot}}
\def\allstar{{AllStar}}
\def\etal{\it et al.\ \rm}
\def\eg{{\frenchspacing e.{\thinspace}g.}}
\def\ie{{\frenchspacing i.{\thinspace}e.}}

\def\spose#1{\hbox to 0pt{#1\hss}}
\def\simlt{\mathrel{\spose{\lower 3pt\hbox{$\mathchar"218$}}
     \raise 2.0pt\hbox{$\mathchar"13C$}}}
\def\simgt{\mathrel{\spose{\lower 3pt\hbox{$\mathchar"218$}}
     \raise 2.0pt\hbox{$\mathchar"13E$}}}

\begin{document}

\title{The MACHO Project: Microlensing Detection Efficiency} 

\author{
      C.~Alcock\altaffilmark{1,2},
    R.A.~Allsman\altaffilmark{3},
    D.R.~Alves\altaffilmark{12},
    T.S.~Axelrod\altaffilmark{4},
    A.C.~Becker\altaffilmark{6},
    D.P.~Bennett\altaffilmark{10,1},
    K.H.~Cook\altaffilmark{1,2},
    A.J.~Drake\altaffilmark{1,4},
    K.C.~Freeman\altaffilmark{4},
      M.~Geha\altaffilmark{1},
      K.~Griest\altaffilmark{2,5},
    M.J.~Lehner\altaffilmark{11},
    S.L.~Marshall\altaffilmark{1,2},
    D.~Minniti\altaffilmark{14},
    C.A.~Nelson\altaffilmark{1,2},
    B.A.~Peterson\altaffilmark{4},
      P.~Popowski\altaffilmark{1},
    M.R.~Pratt\altaffilmark{6},
    P.J.~Quinn\altaffilmark{13},
    C.W.~Stubbs\altaffilmark{2,4,6,9},
      W.~Sutherland\altaffilmark{7},
    A.B.~Tomaney\altaffilmark{6},
      T.~Vandehei\altaffilmark{2,5},
      D.~Welch\altaffilmark{8}
    }
\begin{center}
{\bf (The MACHO Collaboration) }\\
\lastrev   
\end{center}


\altaffiltext{1}{Lawrence Livermore National Laboratory, Livermore, CA 94550\\
    Email: {\tt alcock, adrake, cnelson, dminniti, kcook, mgeha,
    popowski, stuart@igpp.llnl.gov}}

\altaffiltext{2}{Center for Particle Astrophysics,
    University of California, Berkeley, CA 94720}

\altaffiltext{3}{Supercomputing Facility, Australian National University,
    Canberra, ACT 0200, Australia \\
    Email: {\tt robyn@macho.anu.edu.au}}

\altaffiltext{4}{Research School of Astronomy and Astrophysics, Canberra,
    Weston Creek, ACT 2611, Australia
    Email: \tt tsa, kcf, peterson@mso.anu.edu.au}

\altaffiltext{5}{Department of Physics, University of California,
    San Diego, CA 92039\\
    Email: {\tt kgriest, vandehei@astrophys.ucsd.edu }}

\altaffiltext{6}{Departments of Astronomy and Physics,
    University of Washington, Seattle, WA 98195\\
    Email: {\tt becker, mrp, stubbs@astro.washington.edu}}

\altaffiltext{7}{Department of Physics, University of Oxford,
    Oxford OX1 3RH, U.K.\\
    Email: {\tt w.sutherland@physics.ox.ac.uk}}

\altaffiltext{8}{McMaster University, Hamilton, Ontario Canada L8S 4M1\\
    Email: {\tt welch@physics.mcmaster.ca}}

\altaffiltext{9}{Visiting Astronomer, Cerro Tololo Inter-American Observatory}

\altaffiltext{10}{Department of Physics, University of Notre Dame, IN 46556\\
    Email: {\tt bennett@bustard.phys.nd.edu}}

\altaffiltext{11}{Department of Physics, University of Sheffield, Sheffield S3 7RH, UK\\
    Email: {\tt m.lehner@sheffield.ac.uk}}
\altaffiltext{12}{Space Telescope Science Institute, 3700 San Martin Dr.,
    Baltimore, MD 21218\\
    Email: {\tt alves@stsci.edu}}

\altaffiltext{13}{European Southern Observatory, Karl Schwarzchild Str. 2, D-8574 8 G\"{a}rching bel M\"{u}nchen, Germany\\
    Email: \tt pjq@eso.org}

\altaffiltext{14}{Departmento de Astronomia, P. Universidad Cat\'olica, Casilla 10 4, Santiago 22, Chile\\
    Email: \tt dante@astro.puc.cl}

\newpage

\begin{abstract}

The MACHO project is a search for dark matter in the form of massive compact
halo objects (MACHOs).  The project has photometrically monitored tens of 
millions of stars in the Large Magellanic Cloud (LMC), Small Magellanic Cloud
(SMC), and Galactic bulge in search of rare gravitational microlensing events
caused by these otherwise invisible objects.  In 5.7 years of
observations toward the LMC some 13--17 microlensing events have 
been observed by the MACHO survey, allowing powerful statements to be made 
about the nature of the dark population in the halo of our Galaxy.
A critical component of these statements is an accurate determination of the 
survey's detection efficiency.  The detection efficiency is a complicated
function of temporal sampling, stellar crowding (the luminosity function),
image quality, photometry, time-series analysis, and criteria used to
select the microlensing candidates.  Such a complex interdependence is
most naturally solved using a Monte Carlo approach.
Here we describe the details of the Monte Carlo used to calculate the
efficiency presented in the MACHO 5.7-year LMC results.  A similar
calculation was performed for MACHO's 1-year and 2-year results.  Here we
correct several shortcomings of these past determinations, 
including (1) adding fainter source stars (2.5 magnitudes below our faintest
detected ``stars"), (2) an up-to-date luminosity function for the LMC,
(3) better sampling of real images in both stellar density and observing
conditions, (4) an improved scheme for adding artificial microlensing
onto a random sample of real lightcurves, and many other improvements.
The Monte Carlo technique presented here realistically
simulates the negative effects of severe crowding (blending) that
is a fact of microlensing surveys.

\end{abstract}


\def\yrone {A96}  
\def\yrtwo {A97}  
\def\yrfive {A00} 
\setcounter{footnote}{0}

\section{INTRODUCTION}
\label{sec-intro}

Gravitational microlensing has become an important tool for the discovery
and characterization of dark populations.  There are now as many microlensing
groups (\cite{erossmc2,lmc2,AGAPE,MEGA99,udalski-erosII}) 
as there are potential lines of 
sight out of the Galactic halo (LMC, SMC, Galactic bulge, M31).
Surveys toward the LMC have observed an excess in the number of 
microlensing events over what is expected from known populations of stars.
We have recently reported 13--17 microlensing events (\cite{lmc5};
hereafter \yrfive) in 5.7 years of observations toward the LMC and
recently EROSII has reported
two new events that they interpret as limiting the amount of halo dark
matter (\cite{eros-2}), but are consistent with the results of 
\yrfive.  In addition, a number of candidates have been observed towards
the SMC (\cite{smc1,smc2,planetsmc2,erossmc1,erossmc2,smc-binary}).
One natural explanation for this excess rate is a population of
massive compact halo objects of mass $\sim\!0.6 M_{\odot}$ that
contribute $\sim\!20$\% to the mass of our Galaxy's halo.  However,
previously unknown (or underestimated) populations of stellar lenses
(\eg, in an LMC halo) are a distinct possibility. 
In order to make quantitative statements about such a population 
an accurate determination of the survey's microlensing detection 
efficiency is required.  Here we describe the MACHO project's
pipeline for calculating its detection efficiency.  More details   
on the experiment, microlensing terminology, analysis, and interpretation
can be found in the companion paper \yrfive, and
detailed reviews of microlensing in general are given by Roulet \& Mollerach 
1997 and Paczy{\'n}ski 1996.

The detection probability for individual events
depends on many factors, \eg, the 3 event parameters $\Amax$, $\that$,
$\tmax$ (maximum magnification, Einstein-diameter crossing time and
time of peak magnification, respectively), and the unlensed 
stellar magnitude, as well as the
observing strategy and weather conditions.  Such a complicated
dependence is most naturally solved using a Monte Carlo technique.
We may simplify the dependence by averaging over the known
distributions in $\Amax$, $\tmax$, the stellar magnitudes,
and the known time-sampling and weather conditions, to derive our
efficiency as a function only of event timescale, $\Ethat$.

Given an efficiency $\Ethat$ we may compute the observed optical
depth as,

\begin{equation}
\tau_{meas} = {{\pi}\over{4}}{{1}\over{E}}\sum_i{{\that_i}\over{{\cal
E}(\that_i)}},
\label{equation2}
\end{equation}

\noindent
where $E$ is the survey's exposure in object-years,
$\that_i$ is the duration
of event $i$ and $\Ethat$ is the efficiency for detecting 
microlensing with duration $\that$ (see also Alcock \etal 1996 and 1997a;
hereafter \yrone\ and \yrtwo, respectively).
The optical depth is a function of the amount and
distribution of mass along the line of sight and is independent of
the lens masses or velocities.
However, we may extract more information if we
assume a model of the Galactic halo.  With the use of a model,
a likelihood analysis may be performed to determine the most likely
lens mass $m$ and mass fraction $f$ of the halo model.  A model yields the
distribution of event durations $d\Gamma/d\that$ (\cite{explore}), which, 
when combined with the efficiency and integrated over all possible 
durations, predicts the number of events one expects to observe
from such a Galactic halo composed entirely of MACHOs. 

\begin{equation}
N_{exp} = E \int_{0}^{\infty} {\cal E} (\that)\,{d\Gamma \over d\that}\,d\that.
\label{equation3}
\end{equation}

\noindent
An identical method may be outlined for stellar populations as in \yrfive.
A likelihood estimator may be constructed from the
observed events and the model--predicted number of events to determine
simultaneously the most likely lens mass $m$ and mass fraction $f$ of the halo. 
Note the vital role the efficiency plays in connecting the world of models to
that of the survey.

The efficiency is a strong function of the temporal sampling of the survey.
Events with very short durations (or very long durations) are unlikely
to be detected as they easily fall `in-between' observations (or
for long durations, extend through the data window).
A straightforward way of simulating this ``sampling efficiency" while 
retaining realistic behavior of the data is to use a random sample
of real lightcurves from the survey in which to inject artificial 
microlensing.  By randomly generating a number of events and running
the same time-series analysis and selection criteria used in the real
analysis, this Monte Carlo can be used to determine the survey's 
detection efficiency (\cite{lmc1,lmc2,pratt}).  

How does one add microlensing onto a lightcurve?  A simple method
is outlined in \S~\ref{samplinglc} for the case where one assumes
that each lightcurve represents a single resolved star.  In this case
all of the flux represented by the lightcurve is magnified by the
microlensing model $A(t)$ (Equation~\ref{equation6} below).  The rarity of
microlensing, however, demands crowded stellar fields in order to detect
even a handful of events, and this in turn results in the blending\footnote{
Blending is when two or more star's point-spread functions overlap to such
an extent that the photometry code can no longer identify individual stars,
but rather recovers one photometric `blob' or object.}
of stellar light.  The issue now becomes what fraction of the flux in the
lightcurve (or object) is lensed?  Our solution was first described briefly
in \yrone\ (see Pratt 1997 for a more thorough description) 
and is outlined briefly as follows.  The method involves the use of a 
large set of artificial star tests on a sample of MACHO survey images that
are seeded with a color-magnitude diagram of the LMC.  Each artificial star
was added to an entire series of observing conditions and over
15 peak magnifications.  The MACHO survey's dedicated photometry code 
\sodophot\ was then run on the resulting images to determine how each
star behaved over a range of seeing, sky, and magnification.  The resulting
photometry and photometric flags were stored in binary files, which we 
refer to as Photometric Response DataBases (PRDBs).  In this 
``photometric efficiency" technique each artificial star (hence forth
referred to as a Photometric Response Function or PRF) is used as a model
for adding artificial microlensing onto lightcurves.  In most cases only
a fraction of the observed flux in the lightcurve is actually lensed, and
the PRF supplies us this fraction as a function of seeing and sky.  In 
this way both blending and systematic photometry effects are mimicked in
the resulting artificial data.

Blending introduces several problems that can affect microlensing
surveys in serious ways and has been extensively investigated
(\cite{lmc1,wozniak,pratt,hanhubble,han}).  In addition to diluting the true
peak magnification $\Amax$, blending also biases the measured 
durations $\that$ to shorter values, since blended events spend less time
above a given threshold.  This biasing of $\that$ is particularly
important as the optical depth estimate is proportional to the
average duration of the events.  Another effect of blending
(which has received less attention, but is equally important) is that 
the survey's exposure in star-years is typically quite a bit larger 
than is estimated by naively counting photometered objects.
One can view this (for a fixed exposure in object-years) as a net increase
in the efficiency which in part balances out a decrease in efficiency
due to blending.  All of these effects must be taken into account 
if one desires an accurate
detection efficiency known at least to the level of, if not better than,
the shot noise inherent in the low number of events seen toward the
Magellanic Clouds.

We make several corrections and improvements to the Monte Carlo presented
in \yrone\ and \yrtwo.  Most notably these are: (1) we add source 
stars to $V=24.5$, more than 2.5 magnitudes
below our faintest detected stars (and 2.5 magnitudes fainter than was
used in \yrone\ and \yrtwo), (2) we use an up-to-date
luminosity function of the LMC, constructed with our ground-based 
photometry for bright stars ($V < 19$) and HST photometry for dim
stars ($V > 19$), (3) we compute luminosity function normalizations
separately for each MACHO field, (4) 10 fields of size $5' \times 5'$
with widely differing stellar density were used to simulate photometry
of artificial stars over an average of 69 different observing conditions
(only 2 fields with $\sim\!20$ observing conditions were used in
\yrone\ and \yrtwo), (5) an improved scheme for adding artificial
microlensing onto real lightcurves, and (6) we describe a robust method
of statistically correcting for the $\that$ bias that can be used
to estimate the optical depth.

In \S~\ref{sec-sodo} we briefly review the MACHO telescope,
photometry system and time-series analysis in order to introduce
some MACHO specific vocabulary used throughout the rest of the paper.
\S~\ref{sec-sampling} outlines the sampling efficiency approach
as a primer to the photometric efficiency discussion in
\S~\ref{sec-photometric}.  The results of the photometric
efficiency analysis is reserved for \S~\ref{sec-results}, and we
conclude in \S~\ref{sec-summary}.

\section{THE MACHO TELESCOPE, PHOTOMETRY SYSTEM, AND TIME-SERIES ANALYSIS}
\label{sec-sodo}

The MACHO project has full-time use of the $1.27$m telescope at
Mount Stromlo Observatory, Australia, for a period of 8 years
starting July 1992.  The telescope was re-commissioned especially for this
project, and a computer-controlled pointing and drive system was 
installed.  A system of corrective optics has been installed near the
prime focus, giving a focal reduction to $f/3.9$ with a $1$ degree
diameter field of view.  A dichroic beam splitter and filters
provide simultaneous images in two passbands, a ``red'' band 
($\sim\!590-780$ nm) and a ``blue'' band ($\sim\!450-590$ nm).
Two large CCD cameras are employed at the two foci; each 
contains a $2\!\times\!2$ mosaic of $2048\!\times\!2048$ pixel
Loral CCD imagers.  The pixel size is 
$15\mu\!m$, which corresponds to $0.''63$ on the sky, giving a sky 
coverage of $0.7 \times 0.7$ square degrees. 
Details of the camera system are given by Stubbs \etal (1993) and
Marshall \etal (1994) while details of the telescope can be found
in Hart \etal (1996).

The survey's photometry code \sodophot\ is a PSF fitting algorithm based on 
$\dophot$\footnote{\sodophot\ = Son Of \dophot} (\cite{dophot}).
It runs in two different modes.  The first mode, a 
template generation mode, is designed to run on a pair of red and blue  
images or ``chunks." (Each CCD is broken up into 16 roughly $5'\!\times\!5'$
somewhat overlapping regions called ``chunks" for the purpose of 
photometric reductions.)  The red
chunk is reduced first in a manner similar to a standard \dophot \
reduction.  Next the blue chunk is reduced using the results from 
the red chunk to warm-start the reduction, which improves the star 
matching between colors.  This produces a set of pre-templates
for this chunk pair.  Once a set of pre-templates for an entire CCD
(16 chunks) has been generated the final templates are constructed
by enlarging the pre-templates and including stars that lie as
much as $\sim\!90$ pixels outside the nominal chunk boundaries.  This 
enlargement lessens negative effects on the routine photometry due 
to incorrect telescope pointing.  The final set of template files
contains a list of detected objects, their positions (corrected to
an airmass of one) 
and their template magnitudes.  A set of template files need be generated 
only once and is created using images with better than average seeing and
dark sky conditions.

All other images are processed in ``routine" mode, which proceeds as 
follows.  The appropriate template file is used to warm-start the image 
by first locating and matching 50 bright unsaturated stars.  These
stars are used to determine a seven parameter analytic fit to the PSF,
a coordinate transformation, and a photometric zero
point relative to the template.  Then all the template stars are
subtracted from the image using the model PSF and coordinate 
transformation; the variance estimate for each pixel is adjusted to
allow for errors in subtraction.  Next, photometric fitting is
carried out for each star in descending order of brightness, by adding
the analytic model of the star back to the subtracted frame and fitting
a two-parameter fit to the star's flux and sky background, with pixels
weighted by inverse variance.  The model PSF and computed position
of the star are kept fixed.  When a star is found to vary significantly
from its template magnitude, it and its neighbors undergo a second
iteration of fitting. 
For each star the estimated magnitude and error are determined, along
with six other parameters (``flags") measuring (1) the object ``type"
(single/blended, etc.), (2) the $\chi^{2}$ of the PSF fit, (3) the
``crowding," \ie, the amount of flux contributed from nearby 
stars, (4) the weighted fractions of the PSF masked due to cosmic 
rays, (5) the weighted fractions of the PSF masked due to bad
pixels and (6) the fit sky value. 

The set of photometric data points for each detected object are rearranged
into a time series and combined with other relevant information (including
seeing, average sky brightness, airmass, etc.) into portable binary files
called sodsets.  The sodsets are in turn passed into our time-series analysis
code to search for variable objects and microlensing candidates (see
\yrfive\ for more details of the analysis).  In brief the analysis
code calculates for each lightcurve a set of variability statistics, 
average magnitudes, error bars, crowding, etc., and for lightcurves 
deemed interesting (level-1 events) a five parameter fit to microlensing
is applied, where the parameters are the unmagnified red and 
blue fluxes, the peak magnification $\Amax$, the time of peak magnification
$\tmax$, and the Einstein-diameter crossing time $\that$.  Many statistics
describing the significance of the fit are also stored for the level-1 events.

From the complete set of statistics we designed criteria that select out
microlensing candidates from a wide background of noise induced bumps and
variable stars.  Two different sets of selection criteria (criteria set A and B) 
were used in \yrfive, and we refer the interested reader to
this paper for full details of the two selection criteria sets, as we give
only a qualitative `feel' for the criteria here. 
Criteria set A superficially resembles the criteria used in \yrtwo\ and was
designed to be somewhat restrictive in the sense that only 
good quality, high signal-to-noise (S/N) events were selected.  The
criteria are fairly tight and rely strongly on S/N statistics that 
are somewhat microlensing-shape specific.  Only 13 candidates in the
first 5.7 
years of LMC data passed criteria set A.  In contrast criteria set B was
designed to be inclusive in the sense that any event with a flat 
baseline and with one significant and unique bump was included.  However,
criteria set B was somewhat vulnerable to variable stars that exhibit
constant baselines for long periods of time with only one outburst,
such as exhibited by supernovae, and possibly cataclysmic variables
and nova (see \yrfive\ for a full discussion on how these interlopers
were treated).  Some 17 candidates in the first 5.7 years of LMC
data passed selection criteria set B.

The relative looseness of criteria set B over A was due to a set of
new statistics that allowed us to more accurately characterize and remove 
the periodic and quasi-peridoic variable star populations.  With a major
source of background removed the S/N level could be lowered.
One of the main overall differences in the two sets of selection criteria 
is that criteria set A is less sensitive on the whole to moderately blended
and highly blended events as compared to criteria set B.
Also, criteria set B relies less on the microlensing-shape information
and is less likely to be missing exotic forms of microlensing such as 
parallax and binary events.
Although criteria set B has overall higher detection efficiency than
criteria set A the number of detected events compensates and the 
results presented of \yrfive\ are fairly insensitive 
to the choice of cuts.  In what follows we will present both sets of
selection criteria and discuss their differences.

\section{THE SAMPLING EFFICIENCY}
\label{sec-sampling}

\subsection{The One Percent Database of Lightcurves}
\label{onepercent}

A key element in the efficiency determination is the $1\%$ database
of lightcurves.  This database represents an unbiased random selection
of $1\%$ of the lightcurves contained in the MACHO project's LMC survey.
It is unbiased with respect to observed magnitude,
color, sampling rate, spatial distribution, and general data quality.
The $1\%$ database contains $118,645$ lightcurves from 
the top 30 LMC fields\footnote{See \yrfive\ for a complete list
of the 30 LMC fields along with RA and dec centers.} and occupies $2.7$
Gbytes of disk space.  The 
lightcurves begin on day 200 (${= \rm JD}-2,448,623.5$; 1992 July 19) and
end on day 2267 (1998 March 17)
which makes the data window 5.7 years long. 
The mean number of exposures per field is 719, with a range from 180 to 1338.
Approximately $13\%$ of the lightcurves have 
information in only one passband (red or blue).  Of these $\sim\!60\%$ 
have only red lightcurves and $\sim\!40\%$ have only blue lightcurves. 
There is a fair amount of field overlap among the top 30 fields
with about $\sim\!20\%$ of lightcurves duplicated across fields.

\subsection{Modifying Lightcurves: Sampling Efficiency}
\label{samplinglc}

Here we assume that each lightcurve in the $1\%$ database is a single
resolved star, and that our photometry code is ``perfect" in recovering 
flux.  We further assume that the measurement errors involved are dominated 
by photon shot noise.  Given a lightcurve we proceed as follows: (1) first a
robust mean magnitude $\overline{m}$ for each bandpass is computed. (2) Then
a set of event parameters $\umin$, $\tmax$, and $\that$ is generated where, 
(a) $\umin$ is chosen as a uniform deviate from zero to the experimental 
threshold $u_{T} = 1.0$ ($\Amax = 1.34$), (b) 
$\tmax$ is chosen as a uniform deviate from slightly before the beginning of the 
observations to slightly after (day 190 to 2277), and (c) $\that$ is
chosen as a uniform base 10 logarithmic deviate with a duration of
1 day up to 2000 days. (3) For each point on the lightcurve  
we compute the ``scatter" from the mean in flux units as $\Delta\!f = 
f - \overline{f}$, where $\overline{f}$ is a robust mean magnitude expressed
in flux units, \ie, $\overline{m} = -2.5 log(\overline{f})$.
This scatter is assumed to be due entirely to photometric 
error rather than intrinsic variation in the source (indeed, only
$\sim\!1-2\%$ of all LMC lightcurves show signs of intrinsic variability).
(4) The flux $f$
for this point is then ``magnified" $f \rightarrow A(u(t)) \overline{f}
+ \Delta\!f$,
or expressed equivalently in magnitudes, $m \rightarrow \overline{m}
- 2.5 log{(A(u(t)) - 1 + 10^{(\overline{m} - m)/2.5})}$. (5) The flux error is 
modified as $\sigma_{f} \rightarrow \sigma_{f} \sqrt{A(u(t))}$, which
translates into magnitude space as $\sigma \rightarrow \sigma/\sqrt{A(u(t))}$.
Here $t$ is the time of the observation being modified and $u(t)$ is

\begin{equation}
u(t) = \left[ u_{\rm min}^2 + \left({2 (t-t_0 ) / \that}\right)^2
\right]^{1/2},
\label{equation5}
\end{equation}

\noindent
with

\begin{equation}
A(u) = {{u^2 + 2}\over{u\sqrt{u^2 + 4}}}.
\label{equation6}
\end{equation}

\noindent
This is performed for all points on the lightcurve and for both
bandpasses.  We leave the remaining photometric flags ($\chi^{2}_{psf}$,
crowding, etc.) unchanged.

\subsection{Calculation of the Efficiency}
\label{samplingcalculation}

Using the technique of \S~\ref{samplinglc} to modify lightcurves
we proceed to add an event onto each lightcurve in the $1\%$ database.
One complete loop through the $1\%$ database we refer
to as a pass.  For each ``event" we save information such as,
(1) the lightcurve id number (field.tile.sequence), (2) a pass number, (3)
a robust magnitude $V_{\rm{obj}}$ and color ${V-R}_{\rm{obj}}$ of the lightcurve, and 
(4) the input event parameters $\Amax$, $\tmax$, and $\that$.
This bookkeeping information will later be matched with the output of the
time-series analysis and used to calculate the efficiency.

Once we have created a complete pass and stored the bookkeeping information
we next run the time-series analysis on the artificial lightcurves 
(\S~\ref{sec-sodo}).  The resulting output statistics and fits 
are written into a file and then matched on a lightcurve by 
lightcurve basis with the bookkeeping files that contain the input statistics.
Hereafter all parameters labeled with a prime refer to measured
parameters (\eg, $\rAmax$, $\rtmax$, and $\rthat$) and un-primed parameters
refer to input or true parameter values.
In addition, both criteria sets A and B described 
in \S~\ref{sec-sodo} are applied to the resulting statistics,
as in \yrfive, and the results stored as an integer: zero = failed the
criteria set,
one = passed the criteria set.  This creates a single datacube file containing
input statistics, output statistics, and `cut' integers.
In order to generate enough events to ensure adequate statistics we
ran 5 passes through the $1\%$ database.  Though the same lightcurves
were used 5 times each, completely different event parameters were
generated for each pass.  Because we stored a unique pass number in the 
datacube file we simply concatenated all 5 datacube passes into one 
large datacube file for ease of use.

With this datacube it is straightforward to calculate the survey's
sampling efficiency.  Since the artificial events are added with a uniform
distribution in $\umin$ and $\tmax$, integration over these
variables~\footnote{We ignore a small modulation ($\sim5\%$) in the event rate
due to the motion of the earth around the sun.} is not required. 
Furthermore, since our assumption in \S~\ref{samplinglc} is that 
each lightcurve represents a single resolved star, there is also no 
need to integrate over luminosity (since, under this assumption, the
$1\%$ database fairly samples the survey's luminosity distribution).  We need
only bin the data in $\that$ and simply count the number of recovered 
events versus the number of added events.  The efficiency is simply
$\Ethat = {N_{rec}(\that)/N_{add}(\that)}$. 

\subsection{Results of the Sampling Efficiency}
\label{samplingresults}

In Figure~\ref{fig-samp-eff} we compare the sampling efficiency
derived from the 5.7-year data set for both criteria set A and B with
the previous two data sets, \yrone\ and \yrtwo.  The general behavior of the
efficiency with $\that$ is easy to understand.  The gradual fall in
efficiency as $\that$ decreases is caused by a combination of
a typical sampling rate of 1-2 days and large gaps of 3-50 days that
exist due to bad weather and other telescope down time.
The sharp drop at $\sim\!600$ days is due to an explicit cut requiring 
$\that < 600$ days (affecting both criteria set).  The large difference in
the efficiencies for the two selection criteria is due to a combination 
of (1) the different values used for the $\Amax$ cut (criteria set A uses 
$\Amax > 1.49$ while criteria set B uses $\Amax > 1.34$), (2) criteria set B 
uses fewer S/N statistics, most of which are looser, (3) differences
in variable star cuts, and (4) various cuts on the minimum quality of
the lightcurve (minimum number of points in the red and blue passbands,
maximum crowding, etc.).

The most striking difference between the previous two data sets and 
the 5.7-year set is the much higher efficiency at long durations.
Much of this difference is just a reflection of the longer baseline and
an increase in the allowed duration ($\that < 600$ days).
However, a significant amount of
the difference lies in a quirk of the \yrtwo\ data set. 
In the \yrtwo\ data six of the densest 
fields\footnote{These fields are 1, 7, 9, 77, 78, and 79} had their lightcurves
cut in half, roughly, due to an early generation of templates used to reduce the
photometry for these fields.  These early templates made use of a different
star naming convention than is currently used.  Roughly halfway through the
\yrtwo\ data set the current generation of templates was implemented on 
these six fields.  Since there was no straightforward way to cross 
reference the ID's of the stars in each different convention (indeed,
the map was not even one-to-one), and redoing the photometry was too
prohibitive at the time,
it was decided to consider as separate lightcurves the stars in these 
six fields before the upgrade and after the upgrade and to analysis 
them as such.
This was handled in a self-consistent manner and more details can be found 
in \yrtwo.  The primary effect of this was a lowered
efficiency for long duration events, due primarily to (1) an effective
cut on event peak $\tmax$ where the fields were split and (2) the required
minimum of 40 baseline points outside $\pm\,2\,\that$ of the peak of 
the event that was used in the \yrtwo\ selection criteria.
The problem did not exist in the \yrone\ data, thus the rather close
behavior between \yrone\ and \yrtwo\ for long durations,
even though the later had twice the coverage.  The \yrfive\
data has redone all its photometry in these six fields with the current
generation of templates and does not suffer from this problem.

It is also worth noting that in the $\that$ range 1-40 days the \yrfive\
sampling efficiency (for both criteria) is systematically smaller than
the \yrone\ and \yrtwo\ results.  In part this is due to differences 
in the selection criteria, but is also a product of 
including an additional 8 less densely sampled fields into the 
year \yrfive\ analysis.  Though this tends to reduce our efficiency 
it also increases our exposure.  Also note that 
criteria set B is always above criteria set A except for durations less than
10 days.  The slightly lower efficiency of criteria set B in this range
is due to the larger number of required points in the peak 
$N_{peak}\!\geq\!10$ (criteria set A only required 7 or more $2\sigma$ points;
see \yrfive).

In addition to $\Ethat$ it is also interesting to investigate the efficiency 
as a function of other parameters, such as magnitude, impact parameter and 
stellar density.  Also of considerable interest
is the $\that$ bias mentioned in the introduction.  However, we hold off
discussing these until after we have introduced the photometric efficiency
as this scheme more realistically models the systematics of blending and 
the survey as a whole.

\section{THE PHOTOMETRIC EFFICIENCY}
\label{sec-photometric}

In the introduction some of the potential problems associated with
the effects of blending were discussed.  Blending causes systematic
underestimation of both $\Amax$ and $\that$, but also increases the 
number of stars to which the survey is sensitive.
Clearly none of these effects are taken into account by the sampling 
efficiency described in \S~\ref{sec-sampling}.  To correct for this
deficiency, and to be able to generate realistic microlensing events,
we make use of a large database of artificial star photometry as first
described in \yrone\ and \yrtwo.  We have developed techniques that allow
us to use this artificial star photometry to inject synthetic microlensing
onto $1\%$ database lightcurves, replacing the techniques described in
\S~\ref{samplinglc}.  The new techniques allow us to empirically account for 
the blending problems, as well as other systematic photometry effects unique
to \sodophot\, such as various correlations with seeing, underestimation of
error bars and other systematic changes in the photometry flags.

Our method involves a large set of artificial star tests on a sample of the 
MACHO survey's images that are seeded with a color-magnitude diagram of the LMC.
A large set of real images that fairly sample the survey's distributions in 
observing conditions (seeing and sky) were used to create the database of input
photometry versus output photometry for the sample of artificial stars 
and organized into a PRDB.  The PRDB consists 
of $\sim\!60,000$ individual PRFs that
represent how a star behaves as part of a blended object.  Each star was added
to images over a range of observing conditions and over 15
peak magnifications.  \sodophot\ was run on the resulting images to determine 
how each star behaved over the range of seeing, sky, and magnification. 
The resulting photometry and photometric flags were stored in the PRDBs.
Each PRF represents a rule for how a lightcurve would respond to the 
addition of flux over a wide variety of observing conditions.
This can be represented as,

\begin{equation}
A_{\rm rec} = A_{\rm rec}(A, V, R, V_{\rm obj}, R_{\rm obj}, seeing, sky),
\label{equation8}
\end{equation}

\noindent
where $A_{\rm rec}$ is the recovered magnification, given an 
input magnification $A$, stellar magnitudes $V$ and $R$, object magnitudes
$V_{\rm obj}$ and $R_{\rm obj}$, and characterized by an observing 
condition ($seeing$ and $sky$).

To implement the photometric efficiency properly it is important to know
the underlying luminosity function (LF) to some limiting magnitude
(in our case about $V\sim\!24$).  We discuss our determination of the 
LF in \S~\ref{sec-cmd}.  Next, in
\S~\ref{sec-prdb} we describe how the PRDBs are generated, including
the images, point-spread functions, coordinate and photometric 
transformations.  In \S~\ref{sec-prf} we show some example 
PRFs and discuss general behaviors exhibited by the PRFs.
Discussion of the photometric efficiency (hereafter referred to as
just the efficiency) results is left for \S~\ref{sec-results}.

\subsection{The LMC Luminosity Function and Color-Magnitude Diagram}
\label{sec-cmd}

It is important to know the LF of stars in the LMC 
(and to a lesser extent the color-magnitude diagram) in order to 
accurately estimate the efficiency.  This is because the LF,
along with stellar density, dictates how much stellar PSF overlap on 
the sky there is as a function of magnitude and thus how much 
blending one expects.  Also, since the survey's exposure is greater
than the number of photometered objects suggests, and this can increase the 
sensitivity of the survey to microlensing, we need a fairly
accurate estimate of the number of stars in our fields.

\subsubsection{The Shape of the LMC's Luminosity Function}
\label{sec-lfshape}

The MACHO fields typically go incomplete at magnitudes greater than
$V\sim\!20$ but can be complete to $V\sim\!21$.  Because we desire 
to know the LF to at least $V\sim\!24$ we must resort to Hubble Space Telescope (HST) 
Wide Field Planetary Camera 2 (WFPC2) image data.  The MACHO project has
obtained HST WFPC2 data with the F555W and F675W filters for three fields in
the LMC bar.  These are located in the MACHO fields 2, 11, and 13.  For
each field, we have obtained
``shorts" (3-4 $\times$ 30 sec exposures) and ``longs" (3-4 $\times$ 400-500
sec exposures), in both filters.  More details of the HST data reduction
and analysis are contained in Alcock \etal (1999b).  In addition we have 
obtained the reduced HST WFPC2 data from 6 LMC bar fields obtained and
reduced by Olsen 1999 (kindly provided via private communication).  These 6 
fields are in the F555W and F814W filters and have similar ``shorts" and 
``longs" giving completeness limits similar to our HST data.
Olsen's PC was positioned on old LMC globular clusters and so we discarded
all PC data and, in addition, must be careful of a small contamination from
cluster stars that extend into the edges of the WF CCDs.  Details of the 
Olsen HST reductions can be found in Olsen \etal 1998 and Olsen 1999.

In Figure~\ref{fig-macho-hst-lf} is plotted the V band (F555W filter) 
LF for the 3 MACHO HST fields.  Each field is normalized 
to have the same number of stars in the range $17 < V < 20$\footnote{This
region corresponds to where the HST fields are greater than 95\% complete.},
with the fields of higher stellar density, field~2 and 11, 
normalized to field 13 (lowest density and lowest S/N).  The shape of
the 3 LFs appear consistent with one another from $V\sim\!17$,
where shot noise dominates, to $V\sim\!22.5$, where differences in
stellar density between the 3 fields cause differences in completeness.
Figure~\ref{fig-olsen-hst-lf} is similar to Figure~\ref{fig-macho-hst-lf}
but shows Olsen's 6 fields normalized in a similar fashion to the lowest 
S/N field.  Again the shapes of the 6 LFs appear consistent
with one another between the shot noise on the bright end and the differing
completeness on the dim end.

We can quantify the above `by eye' assessment by computing a $\chi^{2}$ 
between fields along with the associated probability of obtaining a value
of $\chi^{2}/\rm{dof}$ worse than the measured value.
We find no significant difference between the 3 MACHO LFs with 
relative measured $\chi^{2}/\rm{dof}$ of 1.065, 1.253 and 1.242 for
the combinations 2-11, 2-13, and 11-13.  These correspond to probabilities
of $p = 0.35$, $0.11$, and $0.12$, respectively, that the value
$\chi^{2}/\rm{dof}$ could be worse 
than the measured value.  Of the 6 Olsen HST fields one stands
out as anomalous in its $\chi^{2}/\rm{dof}$ value.  The LF of 
NGC~1916 is the most discrepant but appears to be so due to 
heavy differential reddening 
(\cite{olsen1,olsen2}) with relative $\chi^{2}/\rm{dof}$ ranging
from a low of
1.589 to a high of 3.777, corresponding to a range of probabilities
$p = 5.3\ee{-3}$ to $1.1\ee{-17}$.  We thus discard this field from 
further analysis.  The remaining 5 Olsen LFs are consistent
with one another, with relative $\chi^{2}/\rm{dof}$ ranging from a low of
1.118 to a high of 1.891, corresponding to a range of probabilities
$p = 0.26$ to $1.6\ee{-4}$.  The slightly worse probabilities 
seen in the Olsen fields are likely due to contamination from the
globular clusters spilling over into the WF CCDs.

In Figure~\ref{fig-olsen-macho-f13-lf} we plot MACHO's combined
LF (3 fields) and Olsen's combined LF (5 fields),
normalized to the MACHO combined LF in the range $17 < V < 20$. 
The two LFs are consistent with one another in the range $17 < V < 22$
($\chi^{2}/\rm{dof} = 1.215$, $p = 0.17$).  Also plotted is MACHO's 
ground-based LF for field 13 ($\sim\!350,000$ objects) which
is one of our most complete, lowest reddened fields, with excellent
template seeing and sky (we have allowed for a small offset of $\sim\,0.1$
in $V$, see below).  The shape of the ground-based LF in the range
$17 < V < 21$ is good in comparison with the MACHO combined HST LF
($\chi^{2}/\rm{dof} = 1.295$, $p = 0.10$).  Given the apparent
lack of any significant difference in
the shapes of the LFs for these 9 LMC bar fields we have chosen to 
combine 8 of them (excluding NGC 1916 because of high reddening) to form a
LF with a fairly high S/N on the dim end (to $V\sim\!22.5$) and to 
splice this together with the bright end ($V < 20$) of the MACHO
ground-based LF.  In this way we create a high S/N LF that is complete
in the range $15 < V < 22.5$.

There is still a question of the completeness of the HST LFs for
magnitudes greater than $V\sim\!22.5$.  We have performed artificial
stars tests on our 3 MACHO HST fields and used 
the resulting completeness curves to correct the 
MACHO LF to $V\sim\!24.5$.  The completeness tests probably
represent a slight underestimate of the true LF.
Because of the potential uncertainty in the shape of the LF for dim stars
we have opted to create two different LFs in which to test for 
systematics.  These are shown in panel (a) of Figure~\ref{fig_eff_per_bin}.
The first LF (hereafter $LF_{1}$) has been extended from $V\sim\!22$ to
greater magnitudes using a power law with a slope of 0.415 (\cite{calib})
derived from a linear regression fit between $19.5 < V < 22$.  This
LF probably represents an overestimate of the number of faint stars.
The second LF (hereafter $LF_{2}$) uses the completeness corrected 
MACHO LF as an estimate of the shape for dim stars.  The shape of 
the true LF is likely to lie somewhere in between these two, but is probably 
closer to the completeness corrected $LF_{2}$.  We choose to use both
LFs in order to estimate any systematic error induced by such an
uncertainty.

\subsubsection{Normalization to MACHO Fields}
\label{sec-lfnorm}

With the shape of the LF determined\footnote{We warn the reader that
the LF in the bar appears to be `universal', and we use this simplifying
assumption for the present results.  The HST fields used here only
represent the LF in the LMC's bar, while the LMC's disk LF has been
shown to be somewhat different in that it lacks an intermediate age
population (\cite{geha,olsen2}).  Our outer most fields 53, 55 and 57
are farther out in the disk and could thus be in error, though we note
that NGC 1754 is also in the disk and, although it shows a different
star formation history than the bar, apparently its LF is still fairly
similar to the bar fields (\cite{olsen2}).} it only remains to find a
normalization, that is,
a number relating these two LFs to the total number of stars.  This is 
most convenient to do on a field by field basis as the stellar density 
changes quite rapidly across our fields (and also inside our fields, but
we average over this).  A normalization is calculated for each field using
the ground-based LF for that field and the `universal' LFs shown in panel (a)
of Figure~\ref{fig_eff_per_bin}.  This is illustrated in 
Figure~\ref{fig-example-norm-lf} where four MACHO fields of widely
differing stellar density are shown.

First, we allow a slight offset in V magnitude 
that is field dependent and typically varies from -0.14 mag to 0.40 mag
(Table~\ref{tab-norms}).
This slight offset is due to a combination of three effects, (1)
extinction (patchy, even on the scale
of our fields), (2) the tilt of the LMC's disk (maximum effect $\pm\,0.07$
in our 30 fields), and (3) fields with poorly calibrated photometry.
We note that the offsets
are not correlated with seeing, sky, airmass or stellar density of the
template images, but are strongly correlated with the color of the LMC 
clump, indicating that differences in extinction are the 
dominate cause.  The
offsets are derived by requiring the peaks of the clump and the tips of
the giant branch to line up in $V$ with the `universal' LFs.  The effect
of extinction is to lower the number of source 
stars to which the survey is sensitive.  Only 22 
LMC fields are well-calibrated (\cite{calib}) and these typically have the
smallest offsets, while 8 LMC fields have only approximate
calibrations and typically have large positive offsets.  Once the 
offsets are applied we next require the number stars in the `universal'
LFs in the range $17.5 < V < 18.5$\footnote{This
region corresponds to where the MACHO ground-based fields are greater 
than 95\% complete.}
to match the number of objects in the ground-based LF in the same range.
The two shifts typically align the 
`universal' LF to the observed LF quite well, both inside and outside
the calibration bin, as seen in Figure~\ref{fig-example-norm-lf}.

Figure~\ref{fig-so22-density} displays the normalizations for each of the
30 MACHO fields derived in the manner described above.  Our choice of units
for the normalization is the ratio of the number of stars to the number of
objects brighter than magnitude $V$. In this case $V = 22$ is convenient
as there are few objects dimmer than this limit.  The $S/O(V<22)$ 
ratios are plotted versus the average density (in $\rm{objects}/\Box'$) 
for each field
as solid circles and labeled with the field number.  The normalizations are
also supplied in Table~\ref{tab-norms}.  Note the general
trend of increasing $S/O(V<22)$ with increasing object density, which 
is simply a reflection of the fact that our fields become less complete
in more crowded fields.  The substantial amount of scatter seen at any 
given density is a reflection of other factors affecting completeness.
For example, $S/O(V<22)$ is strongly correlated with template sky for any
given density, in the sense that fields with high sky tend to be less complete 
(have higher $S/O(V<22)$).  Though $S/O(V<22)$ appears not to be
significantly correlated with template seeing this is not surprising as
the range in seeing for our templates is not nearly as large as the 
range in sky. 

Using our 3 HST fields we may make a direct determination of $S/O(V<22)$ by
simply counting the MACHO objects that lie inside the WF CCDs.  These are
shown in Figure~\ref{fig-so22-density} as solid triangles.
We have connected points from the same fields with a solid line.  Note that
the density shown for the field $S/O(V<22)$ normalizations is an average
over the whole field, while for the HST $S/O(V<22)$ normalizations it is
only over the WF CCDs.  In all three cases the HST frames are in higher 
than average density regions of the MACHO fields, resulting in higher 
$S/O(V<22)$ ratios.  The overall agreement between $S/O(V<22)$
as determined using the individual field normalizations and as determined 
directly with HST is reassuring. 

We have also created three synthetic images as an additional check on the 
normalizations and to look for possible biases due to blending.
Briefly, the synthetic images are $5'\times5'$ in size
and were generated using empirical PSFs (see \S~\ref{sec-prdb}) derived 
from the template images.  A LF similar to $LF_{1}$ above was used to
add $\sim\!500,000$ stars down to $V = 24$ uniformly over the image.
A uniform sky with Poisson noise was added to match the ADU/pixel
distribution and then \sodophot\ was run in template generation mode on these
synthetic images.  The resulting number of recovered objects and sky were 
compared with the real image values simulated.  Several 
iterations over the number of added stars versus added uniform sky were 
required to match the observed number of objects and sky level.
The $S/O(V<22)$ ratio for each of these synthetic images is shown as an
open triangle in Figure~\ref{fig-so22-density} and is connected by a
solid line to its corresponding field average.
The densities of the three synthetic images are in general different
from the corresponding field averages, based on where the chunk lies
in the field.  Note the very similar behavior of the synthetically 
derived $S/O(V<22)$ compared with the field average $S/O(V<22)$ values
and the HST derived $S/O(V<22)$.  In all three cases the higher density 
had the higher $S/O(V<22)$ ratio.  Also note the much larger increase
in $S/O(V<22)$ for the higher densities.  This suggests a saturation point
is reached in our densest fields whereby the addition of more stars
results in a dwindling increase in the number of detected
objects as the stars pile on top of one another in the image.  We see
a similar behavior in the synthetic images where to recover a
given number of objects one must add a proportionally larger number of 
stars.

Using the synthetic images we may also estimate the size of a possible bias
in the normalizations due to blending.  It would be desirable for our 
normalization bin ($17.5 < V < 18.5$) to contain an equal number of 
photometered objects and real stars: \ie, $N_{\rm{objects}}(17.5 < V < 18.5)/
N_{\rm{stars}}(17.5 < V < 18.5) = 1.0$.  Our synthetic images have values of
$1.04\!\pm\!0.08$, $1.04\!\pm\!0.05$, and $1.18\!\pm\!0.04$,
where the errors are 1$\sigma$ Poisson errors.  They all appear to 
have slightly more photometered objects than real stars in the 
normalization bin, though two are entirely consistent with 1.0.  Only
the highest density image has a value that is significantly larger
than 1.0 (but by 4.5$\sigma$).  This has the
potential of biasing our derived values of $S/O(V<22)$ to larger values.
However, of the three synthetic images two have extreme densities (as 
these were chosen to look for this effect) and are not 
representative of the average density in our fields (see
Figure~\ref{fig-so22-density}).  Because we expect the biases to be 
strongest in our most crowded fields (and to a lesser extent worse seeing
template images) the overall bias is certainly much less than the 18\% 
seen in the extreme artificial chunk, and probably even less than the
straight average of $\sim\!9$\%.  Another estimate\footnote{This 
second estimate uses a simple physical model of blending that fits 
a model through the three values of 
$N_{\rm{objects}}(17.5<V<18.5)/N_{\rm{stars}}(17.5<V<18.5)$ versus
$N_{stars}(17.5<V<18.5) (seeing)^2$ and extrapolates the results to all
30 fields.  A net bias of 6-8\% is estimated.} puts the possible
bias at 6-8\%.  Given the fairly small size of the effect ($<10$\%)
we make no correction for this bias, but include it in our error budget 
(see \S~\ref{sec-errorbudget}).

The errors in $S/O$ (Table~\ref{tab-norms}) were estimated as follows:
(1) we allowed an uncertainty in the offsets of 0.1 mag, based on the
scatter in the offsets between fields, and (2) based on the estimates
of the maximum blending bias discussed above, the normalizations could
be at most 10\% too high and probably only 5\% too low.
We then propagated these two uncertainties through the normalization 
procedure to produce the final errors in the table.
The average error in $S/O$ per field is $\sim\!20\%$. 
The density ($\rm{objects}/\Box'$) weighted mean number of stars to
objects with $V < 24$ is $\overline{S/O(V<24)} = 17.05\!\pm\!3.0$ for $LF_{1}$
and $\overline{S/O(V<24)} = 10.83\!\pm\!1.9$ for $LF_{2}$.
The normalization of each field amounts to correcting the MACHO
nominal exposure in \textit{object-years} to an exposure in
\textit{star-years}.  The uncertainty in the normalization
is moderately large and translates directly into a moderately
large uncertainty in the efficiency as discussed in 
\S~\ref{sec-results}.

\subsubsection{LMC Color-Magnitude Diagram}
\label{sec-lmccmd}

Since we simultaneously image in both a red and a blue passband we require
a color-magnitude diagram (CMD) from which to draw our artificial stars.
Such a CMD was created for precisely this purpose and its construction is
described in detail in Alcock \etal (1999b).  Briefly, we spliced the bright
end ($V < 18.7$) of a CMD derived from $\sim\!9$ million objects 
(\cite{9mcmd}) from our 22 calibrated ground-based fields onto the dim end 
($V > 18.7$) of a CMD derived from our combined 3 HST fields (which is
scaled by relative sky area to the 9M CMD).  The HST CMD was corrected on the
dim end ($V > 22$) using a completeness function that corrected the data to 
a slope of 0.415 in $log(dN/dV)$ versus $V$.  A small amount of
editing was also required to remove isolated high pixels (due to the scaling
up of the lower S/N HST data) and bright foreground stars.  The CMD used
here need not be accurate (though we endeavored to make it so) as its only 
purpose is to draw artificial stars from a realistic distribution in 
color.  We expect no dependence of the efficiency on color
(and see none) and so henceforth discard any color dependency.

\subsection{The Photometric Response DataBase}
\label{sec-prdb}

To create the photometric response database we selected 10 sub-images,
approximately $5'\,\times\,5'$ in size (``chunks"), that span the range
of stellar densities observed in the 30 MACHO fields ($\sim\!93$ to
$\sim\!317$ $\rm{objects}/\Box'$; see Table~\ref{tab-norms}) and that lie 
at a range of distances from the optical center of the bar (from 
10 arcmins to 3.2 degrees).  For each of these 10 chunks we extracted 
from the MACHO archive a range of observing conditions that fairly 
sample the survey's distribution in seeing and sky.  Table~\ref{tab-chunks}
lists some relevant parameters for
each chunk, including MACHO field and chunk id, template observation number,
density in $\rm{objects}/\Box'$, number of observing conditions $N_{cond}$,
number of stars per grid $N_{PRF}/grid$ (see below), and the total
number of useful PRFs.  The mean number of observing conditions for these 10 
chunks is 69, a factor of 3 more than used in \yrone\ and \yrtwo.

In order to add artificial stars to these images we require knowledge of the
point-spread function (PSF) for each image.  We also need a photometric and 
coordinate transformation (CTR) that maps the photometry and position of 
the artificial stars to some reference image, in this case the template images
for each chunk.  We generated empirical PSFs using a PERL script that 
automates the astronomical photometry package \daophot/\allstar\
(\cite{stetson}).  The script iterates multiple times with a series of
simple analytic PSFs searching for stars,
fitting, and subtracting neighbors to candidate PSF stars on each 
iteration while using a progressively more realistic model of the PSF.
The final PSF is a Lorentz-Gaussian analytic model with an empirical 
look-up table of corrections.  PSFs for each of the $\sim\!1600$ chunks 
are generated in this fashion, and are of higher quality than 
the simple modified Gaussian analytic model used by \sodophot.
We visually inspected
a number of PSF subtracted images and were quite satisfied with the
subtraction.  \allstar\ photometry, using these PSFs, was run on all 
$\sim\!1600$ chunks and the results fed into a custom code that searches
for and computes both photometric and coordinate transformations.  The code
is based on the Groth (1986) algorithm and incorporates a color dependent term
in both photometric and coordinate transformations.  The inclusion of color in
the transformations allows for the effects of airmass, differential refraction,
and a known systematic pier side CCD effect in the MACHO data (``blue 
jitter"; see Alcock \etal (1999b)).  From the residuals to the fits we find the 
coordinate transformations are good to $\sim\!0.18$ pixels and the
photometric transformations to $\sim\!0.08$ mag.

Grids of artificial stars are added to each of the 10 chunks over all
$\sim\!69$ observing conditions using the appropriate PSFs and CTRs 
for each observing condition.
The grids contain from 64 to 156 artificial stars (Table~\ref{tab-chunks})
positioned on a pseudo-random spatial grid
such that stars are never closer than $\sim\!21$ arcsecs (or 
approximately $\sim\!7$ FWHM in median seeing) and never closer than
$\sim\!20$ arcsecs to a boundary of the chunk.  This is to avoid missing 
too much data because of telescope pointing errors.  The magnitudes of
the artificial stars, $V$ and $R$, are drawn from the CMD of \S~\ref{sec-lmccmd}
in the range $16.5\!<\!V\!<\!24.5$ and $-0.5\!<\!V-R\!<\!1.5$. 
The CMD is sampled
in a square-root fashion to ensure uniform statistics over the large range in 
luminosity.
Each star in 
the grid is added over a range of 15 peak magnifications $A = 0.8, 1.0,
1.1, 1.2, 1.34, 1.5, 2, 3, 4, 5, 10, 20, 30, 40, 50$.  In each image the
peak magnifications for each star are staggered so as not to add too 
many highly magnified stars to a given
image.  To cover as much of each chunk's image plane as possible, 60 different 
grids are created for each chunk.  This corresponds to covering
$\sim\!30-70\%$ of the image plane (depending on the chunk) and represents 
a statistically significant number of possible photometric conditions in
which a star could reside.

All stars in a grid are first added with a magnification of one to 
the template
image of the chunk and \sodophot\ is run in template generation mode on these
template images to create the template files.  With the template files created,
we next loop through all observing conditions and all possible peak
magnifications,
adding the grid to the corresponding observing condition using the appropriate
PSFs, CTRs, CCD gains, and with a staggered magnification for each star.
\sodophot\ in routine mode is run on each image created and
the results
are organized and stored in PRDB files.  We chose to store all photometry
on the three spatially nearest recovered objects to each artificial star in
order to map out where the added flux goes.  For each grid of stars
a PRDB file with 17.4 Mbytes of photometry is created.  In total the PRDB
contains 10.4 Gbytes of photometry on 196,740 recovered objects.

The total number of artificial stars added to all 10 chunks over the 60 grids
per chunk is 65,580.  Of these 54,981 artificial stars contain enough
reduced data (greater than 50\% in each passbands) to be useful as PRFs.
This is a factor of five more PRFs than was used in 
\yrone\ and \yrtwo.  Approximately 16\% of the PRFs were lost due
to a combination of (1) falling
inside an obliterated region caused by a saturated star ($\sim\!2\%$), (2) 
missing data entirely in either the red ($\sim\!3\%$) or blue ($\sim\!6\%$) 
bandpass, and (3) falling near an obliterated region, CCD defect, or 
chunk edge such that on average more than 50\% of the photometry was 
missing ($\sim\!5\%$).  Since we are only concerned with how \sodophot\
responds to added flux the loss of poorly determined PRFs is not a concern.
The number of single bandpass lightcurves, CCD defects, and other missing
data are preserved in the 1\% database of lightcurves.

\subsection{The Photometric Response Functions}
\label{sec-prf}

Initially we were concerned with how \sodophot\ divides up flux 
between two or more closely spaced stars, and which neighbor is
most `sensitive' to the added flux.  For example, in \yrfive\ two 
lightcurves (events~7a and 7b) are, in fact, the same event.  
Event~7 was bright enough and
in a locally crowded enough region that some of the flux from the primary
(7a) bled into a secondary neighbor (7b) causing a spurious detection.
Although event~7b was removed from the list of microlensing events and had no
adverse effect on the results of \yrfive, it underscores the need to
investigate what effect multiple crowded neighbors have on our
efficiency.  To investigate this we stored the photometry for the three 
spatially nearest objects to each artificial star in the PRDBs.  We
found that in the vast majority of cases ($\sim\!97\%$) 
the nearest neighbor snatched most of the added flux and was the
most `sensitive'.  This is reassuring, both for simplicity's sake and 
because our previous work (\yrone\ and \yrtwo) had implicitly assumed
this behavior.  We illustrate this as follows.

For each neighbor we define a recovered magnification, $A_{rec}(i,j) =
f(i,j)/{\overline{f}}$, where $f(i,j)$ is the flux of the neighbor for 
observing condition $i$ and peak magnification $j$.  Here
$\overline{f}$ is the baseline flux for the neighbor (that is the flux
averaged over all $i$ with $j = 0$ corresponding to $A = 1.0$).  A 
scatter plot of $A_{rec}$ versus $A$ for all $i$ shows an excellent
linear relationship as expected (see below), with a fit slope
$m = dA_{rec}/dA$ that is generally between 0 and 1.  This slope 
corresponds to the average blend fraction.  We constructed
a measure of sensitivity to added flux and computed this measure for each of
the three neighbors to each artificial star.  The measure we used was the
fit slope $m$ divided by a mean relative error bar $\sigma$ for the
baseline, \ie, measure = $m/\sigma$.  This measure ensured that neighbors
with large slopes and small error bars are counted as the most sensitive,
while in the case of two neighbors with equal slopes the neighbor with the
smallest error bars is counted as the most sensitive.  This is likely to
be the case as both selection criteria set A and B make
use of a signal-to-noise cut requiring the magnification to exceed some
multiple of the mean relative error bar (see \S~\ref{sec-sodo} and \yrfive).

For each artificial star, the three nearest neighbors are ranked according
to the measure, and a cut $\Amax > 1.34$ was applied to each neighbor
to ensure that the PRF was not `junk' (here $\Amax$ is the maximum
magnification of the recovered neighbor). 
Only in $\sim\!3$\% of the cases was either the 2nd or 3rd nearest
neighbor more `sensitive' than the nearest neighbor.  We visually
inspected a large number of these cases.  In most cases
there were two neighbors about equally spaced from the artificial star
with neighbor~1 (the closest) dimmer than neighbor~2.  Neighbor~2 was
more `sensitive' for two reasons, (1) being brighter it had the smaller
relative error bar and (2) due to a slight systematic bias in \sodophot\ the
brighter stars, which are reduced first, can pirate flux from the
wings of nearby dim stars.  Since only $\sim\!3$\% of cases are in error
due to this effect, we restrict our attention to only the nearest 
neighbor.

A scatter plot of $A_{rec}$ versus $A$ over all observing
conditions is shown for four PRFs in Figure~\ref{fig-prfs}.  Note the 
excellent linear relationship fit by the solid line.
The dashed line is the simple analytic response function used in the 
sampling efficiency (\S~\ref{samplinglc}) and is plotted here for comparison.
The panels correspond to (a) a unblended PRF ($m = 1.0$),
(b) a somewhat blended PRF ($m = 0.77$), (c) a moderately blended PRF 
($m = 0.65$), and (d) a heavily blended PRF ($m = 0.35$).  The scatter
at each $A$ is composed of $\sim\!69$ observing conditions and 
is typically well correlated with seeing in the sense that worse seeing
induces slightly larger magnifications.  This is easy to
understand, since worse seeing implies the PSFs overlap more and thus more
flux can be contributed to the nearest neighbor by a magnified
star.  A few cases of very high positive correlation and 
even negative correlation exist and correspond to cases where (1)
the artificial star is not directly detected and lies a 
moderate distance from the nearest neighbor so that when it
is magnified it only affects its nearest neighbor in poor seeing
and (2) the artificial star lies almost equidistant between the two nearest
neighbors and in good seeing contributes to the nearest neighbor but
in bad seeing contributes more to the second nearest neighbor.

To investigate how the relative error bars $\sigma$ (error bars
expressed in magnitudes) behave versus $A$ we define an
effective error `de-magnification' as $A_{\sigma}(i,j) = \sigma(i,j)/
\sigma(i,j=0)$, where $\sigma(i,j)$ is $\sigma$ for observing
condition $i$ and peak magnification $j$.  We normalize
the error bar magnification separately for each observing condition $i$
because $\sigma$ is in general highly correlated with seeing
(becoming larger in worse seeing).  Notice that
$A_{\sigma}(i,j)\,\rightarrow\,1.0$ for all $i$ when $j\,\rightarrow\,0$
(that is $A\,\rightarrow\,1.0$).  Figure~\ref{fig-prferrors} is a scatter 
plot of $A_{\sigma}$ versus $A$ for the same set of PRFs as shown 
in Figure~\ref{fig-prfs}.  Over-plotted as a solid line on each panel is 
the purely Poisson behavior used in the sampling efficiency
($A_{\sigma} = 1.0/\sqrt(A)$).  The behaviors 
of the PRF's relative error bars show little resemblance to this
purely Poisson approximation.  Rather the PRFs behave more like 
$A_{\sigma} = 1.0/A$ (dashed line in the figure),
which can be understood as follows.

In our most common case the noise
is dominated by a combination of the Poisson noise in the sky plus extra
Poisson noise added due to neighboring stars which have been subtracted
during the photometry reductions.  These are both independent of
magnification, so the error in flux should not depend on magnification.
That is, the error in linear flux units should not change at all,
while the relative errors should behave as $\sim\!1.0/flux$ or, as in our
case $\sim\!1.0/A$.   This is the pattern we see in our
PRFs.  However, very bright stars typically fall into the purely Poisson 
limit.  Heavily blended PRFs can also approach the Poisson limit, and 
in some cases their relative error bars change very little as the magnified
artificial star is only a small perturbation on the much brighter neighbor. 

The broad range in behavior of the PRFs is illustrated in 
Figure~\ref{fig-slopes}, which is a scatter plot of fit
slope $m = dA_{rec}/dA$ versus magnitude $V$ for
$\sim\!59,000$ PRFs.  The amount and distribution of blending is immediately
apparent in this plot, with the artificial stars being strongly bifurcated
towards either being recovered within $\pm\,10\%$ of their input flux or being
blended by greater than $\sim\!90\%$.  However, there are still a substantial 
number that are recovered at intermediate blend fractions, and as the
artificial stars shown here represents a square-root sampling of the CMD 
(thus under-weighting dim stars; see \S~\ref{sec-prdb}) the corresponding
results for a linear sampling would show proportionally even more blending.  
The left ordinate is the fit slope $m$ and the right ordinate is scaled to the 
maximum recovered magnification that each PRF contains (due to our 
maximum input magnification of 50).  For example, an artificial star
with $m\!=\!0.5$ would contain a maximum recovered magnification of 
$A_{rec}\!=\!25$ in the PRF (with any higher magnifications needing to
be extrapolated beyond this point, see \S~\ref{modifylcsection}). 

Also plotted in Figure~\ref{fig-slopes} is a family of five smooth curves
that correspond to model $m(V)$.  Each curve in the family assumes a 
certain amount of blended flux corresponding to $V_{blend}\!=\!18,\,19,\,
20,\,21$, and $22$ mags in the figure.  For example,
a star blended with 20 mags of additional flux (the center curve) would
fall at $m\!=\!0.7$ if it was $V\!=\!19$, $m\!=\!0.5$ if it was $V\!=\!20$,
and $m\!=\!0.25$ if it was $V\!=\!21$.  Note how the family of curves
brackets the scattered points fairly well.  The solid horizontal line 
labeled $A\!=\!1.75$ illustrates that stars fainter than $V\sim\!24$ 
rarely, if ever, are magnified greater than this limit.  As a consequence,
if one desired to add stars fainter than $V\!\sim\!24$, one would also need 
to add them at magnifications greater than 50.

The PRFs also allow us to model how the photometry flags
(crowding, $\chi^{2}_{PSF}$, fit sky, etc.) are handled by
\sodophot\ under various observing conditions
and peak magnifications.  We briefly summarize the noteworthy effects here.
The crowding parameter, which is a measure of the amount of contaminating
flux from nearby neighbors, not surprisingly, is highly correlated with seeing.
As such we define a multiplicative `magnification' in a fashion identical
to the relative error bars discussed above.  In most cases ($\simgt\!80\%$)
there is little or no variation of crowding with $A$.  However, in a
few cases the crowding parameter increases smoothly with $A$.
The $\chi^{2}_{PSF}$ is also strongly correlated with seeing, but in the
sense that the fit to the PSF is worse in better seeing.  This
anti-correlation is mostly due to the photometered object being made up
of multiple overlapping PSFs, which are smoothed out in bad seeing.
Unlike the crowding parameter, the $\chi^{2}_{PSF}$ is anti-correlated with 
$A$, in the sense that high magnifications result in poor fits to the PSF.
The fit sky parameter is not significantly correlated with either seeing or
sky, and in only a few cases ($\simlt\!10$\%) is the fit sky well correlated 
with $A$, usually in the sense that it is higher for larger peak
magnifications, but not always.
Neither the missing pixel or cosmic ray flags significantly correlate with
seeing, sky, or magnification.  The object type is more complicated,
but fortunately for our purposes is not important.  How the object type
changes with seeing and sky is already properly handled in the 1\% database
and how it might change with magnification is inconsequential
as long as it remains a valid type.
 
\subsection{Modifying Lightcurves: Photometric Efficiency}
\label{modifylcsection}

We now replace the sampling efficiency rules (\S~\ref{samplinglc})
for modifying 1\% database lightcurves with a new set of rules derived
from the PRFs.  The PRFs offer us an empirical set of rules that 
realistically incorporate blending and the many systematic photometry 
effects observed in \sodophot\ that were described in the last section.

To add microlensing onto a lightcurve with the PRFs we must first match
a lightcurve to a PRF.  This is most consistently performed using
only the available measured parameters of the lightcurves, such as magnitude,
color, average crowding, average error, etc.  We have limited ourselves to 
three parameters, $V_{\rm{obj}}$, $V-R_{\rm{obj}}$, and crowding for the match.
A match in average error was not chosen as it is already highly correlated with
magnitude.  Instead we chose to match in crowding for the following reasons,
(1) the average error is also correlated with crowding (independent of 
magnitude), (2) the crowding parameter is a natural measure for
parameterizing the image plane (\ie, regions of high stellar density
versus low stellar density) as there is a strong
correlation ($r = 0.95$) of the average crowding in MACHO
fields with stellar density, and (3) of all the measured parameters,
we believe crowding to be the most likely to be connected with 
blending.  Though we see no strong correlations between blending and
crowding, we do see a bifurcation reminiscent of Figure~\ref{fig-slopes}
with a slight preference for strongly blended events to also be highly 
crowded.

To match our two large databases (1\% database and PRDB database) as
uniformly as possible we use the following scheme.
First, the 1\% database of lightcurves
is sampled uniformly because of the importance of correctly
weighting the temporal sampling and the unbiased nature of the
database.  For each
lightcurve in the 1\% database a robust mean magnitude $V_{\rm{obj}}$, color 
$V-R_{\rm{obj}}$, and crowding $C_{\rm{obj}}$ are computed.  The PRFs are binned
in three dimensions corresponding to their baseline magnitude $V_{PRF}$,
color $V-R_{PRF}$, and crowding $C_{PRF}$, with bins sizes of $\sim\!0.2$
magnitude ($N_{bins}\,=\,28$), $\sim\!0.1$ color ($N_{bins}\,=\,17$), and
$\sim\!30$ crowding \footnote{The crowding parameter ranges from 0 to 255
and is logarithmic in nature.}
($N_{bins}\,=\,7$).  The bin sizes are constant over the most dense regions of 
their respective distributions, but grow slightly in size near the edges 
to accommodate for sparseness.  The mean number of PRFs in each bin is
$\sim\!17$.  However, due to the shape of the CMD $\sim\!51\%$ of these bins
are empty (as they should be) and thus the mean number of PRFs in occupied
bins is $\sim\!34$, with the typical occupied bin containing 12-20 PRFs.
With the large number of PRFs employed it is unlikely ($\sim\,1\%$) that
a lightcurve will encounter an empty bin.  When this does occur we simply
match to the nearest non-empty bin (while holding crowding constant) which
is never more than one bin away.  The PRF bin that best matches the
lightcurve is then selected and a PRF is randomly chosen from this bin.
In this way we sample the PRFs as uniformly as possible.  

Once a PRF has been matched to a lightcurve, event parameters 
$\Amax$, $\tmax$, and $\that$ are generated as in \S~\ref{samplinglc}. 
The event is now added onto the lightcurve using the rules from the PRF.
Each observation on the lightcurve with $A(t) > 1.005$ is 
matched to the closest corresponding observing condition, $i$, in
the PRF.  In practice this amounts to 
minimizing the quantity $\Delta\!s = (\Delta\!seeing)^2 + (\Delta\!sky)^2$ to
determine $i$, where $\Delta\!seeing$ is the difference between the
observation's seeing and the seeing in one of the $\sim\!69$ observing
conditions contained in the PRF, normalized to the maximum range of seeing. 
$\Delta\!sky$ is similarly defined and normalized.

Having determined the observing conditions $i$, a two point linear 
interpolation
is used to compute $A_{rec}$ based on the low point $A_{lo}\,=\,A_{rec}(i,j)$ 
and high point $A_{hi}\,=\,A_{rec}(i,j+1)$ that bracket $A(t)$ (or the lowest 
two points for an extrapolation).  To preserve as much of the intrinsic 
lightcurve scatter as possible a technique similar to that outlined in 
\S~\ref{samplinglc} is employed in magnifying the flux.  That is
each magnitude is modified as 
$m\,\rightarrow\,\overline{m} - 2.5 log{(A_{rec}(u(t))
- 1 + 10^{(\overline{m} - m)/2.5})}$.  The relative error bars
are modified as $\sigma\,\rightarrow\,A_{\sigma} \sigma$,
where $\sigma$ is the relative error bar for this observation
and $A_{\sigma}$ is also computed via a linear interpolation
between $A_{lo}$ and $A_{hi}$ from the tabulated values of
$A_{\sigma}(i,j)$ (\S~\ref{sec-prf}).
This technique preserves the intrinsic size of the lightcurve's
relative error bars and realistically modifies how they respond
with $A$ over the various observing conditions.

The PRFs also allow us to modify the photometry flags recorded by \sodophot.
As discussed in \S~\ref{sec-prf} some of these show clear signs of
systematic behavior with magnification.  We chose to modify (1) the crowding
parameter, which represents the amount of flux contributed from nearby stars,
(2) the $\chi^{2}_{PSF}$, which tends toward worse fits in better seeing and 
high magnification, and (3) the fit sky value.  We do not modify (4) the 
object type flag, (5) the weighted fractions of the PSF masked due to
cosmic rays, and (6) the weighted fractions of the PSF masked due
to bad pixels.

The flags are modified as $Flag\,\rightarrow\,Flag - dFlag$, where $Flag$
is the respective flag (crowding, $\chi^{2}_{PSF}$, or fit sky) and $dFlag$
is a linearly interpolated difference between the PRF's flag value at
$A$ and at $A = 1.0$, computed separately for each observing
condition $i$.  
Because the flags are logarithmic in nature the magnification translates 
into an additive term, much like in the case of magnitudes.  This 
technique preserves the lightcurve's intrinsic flag values while realistically 
altering them as a function of $A$ and observing conditions.

A sample lightcurve modified using the four PRFs displayed in
Figures~\ref{fig-prfs}~and~\ref{fig-prferrors}
is illustrated in Figure~\ref{fig-artif-lc}.  In each panel the 
input event parameters were the same, $\Amax\,=\,2.0$ and $\that\,=\,80.0$
days.  The panels correspond to blend fractions (a) $m = 1.0$, (b)
$m = 0.77$, (c) $m = 0.65$, and (d) $m = 0.35$.  Note the dramatic drop
in magnification $\Amax$ and relative shortening of the duration $\that$ 
as the event becomes more blended.  Also note the chromatic behavior 
exhibited in PRF (c) due to the difference in color between the lensed
star and the blended flux.
We have visually inspected a large number of artificial lightcurves and
compared the modified portions with real events seen towards both the LMC 
and Galactic Bulge.  The comparison of magnification, error bars, and
flags and their correlations with seeing and magnification are all quite
satisfactory.

\section{RESULTS OF THE PHOTOMETRIC EFFICIENCY}
\label{sec-results}

A procedure similar to that outlined in \S~\ref{samplingcalculation}
is used to compute the photometric efficiency, with one important difference.
In \S~\ref{samplingcalculation} the distribution in luminosity of the 
stars was assumed to be the same as that observed by the survey 
(\ie, uncorrected for incompleteness) since all photometered objects
were counted as resolved
stars.  Since we are now adding events onto lightcurves using an underlying
luminosity distribution we need to integrate over this distribution, as
the observed distribution is clearly incorrect (\S~\ref{sec-cmd}).
In principle, one could quantify the efficiency as a function of 
both duration $\that$ and luminosity $V$, and use this directly. 
However, in practice we cannot unambiguously 
determine the unlensed luminosity (at least without additional intensive
follow-up photometry for blending fits or photometry from space).  Since
the LMC's LF is well known 
to $V\!\sim\!24$ we opt to integrate out this variable.

To generate adequate statistics we make 10 passes through the 1\% database
of lightcurves, matching each lightcurve to a PRF, which inturn is used
to modify the lightcurve (\S~\ref{modifylcsection}).  Bookkeeping 
information (lightcurve id, PRF id, $V$, $V-R$, $V_{\rm{obj}}$,
$V-R_{\rm{obj}}$, $\Amax$, $\tmax$, $\that$, etc.) for all 
lightcurves is stored and
matched to the corresponding statistics ($\rAmax$, $\rtmax$,
$\rthat$, etc.) generated by the time-series analysis
(\S~\ref{sec-sodo}).  Selection criteria set A and B are applied to these 
statistics and the results stored as integers in the datacube.

\subsection{Efficiency}
\label{subsec-phoeff}

The datacube is binned in two dimensions, $V$ and $\that$. 
Recall that events are added uniformly in $\tmax$ and $\umin$, and so 
these are averaged over.  We choose 100 bins in $V$ of size 0.1 mag 
in the range $15 < V < 25$, and 24 bins in 
$\that$ that are logarithmically spaced in the
range $1.0 < \that < 2000.0$ days (Table~\ref{tab-eff}).
Our artificial events are added uniformly in $log(\that)$, so 
bins logarithmically spaced insures equal numbers of events in
each $\that$ bin.  For each bin an efficiency is computed: 
$\epsilon(\that,V) = N_{rec}(\that,V)/N_{add}(\that,V)$ where
$N_{rec}(\that,V)$ and $N_{add}(\that,V)$ are the number of events that pass
the selection criteria set and the number of added events in the bin of
$\that$ and $V$, respectively.
It is worth mentioning that because $\epsilon(\that,V)$ is computed
separately in each $V$ bin, this function is independent of the
CMD used to seed the artificial star tests.
From 10 passes through the 1\% database,
$\sim\!1.2$ million artificial events are generated and of these 
only $\sim\!43,000$ pass criteria set A and $\sim\!52,000$ pass criteria set B.
The mean number of added events per bin is 500 and the mean number of
recovered events per bin is 18 (criteria set A). 

The function $\epsilon(\that,V)$ for criteria set B is shown in 
Figure~\ref{fig_eff_per_bin} as a contour plot in panel (b).  The contours
correspond to efficiencies of 0.001, 0.01, 0.1, 0.2, 0.3, 0.4, and 0.5.  
The gross behavior with $\that$ and $V$ is apparent, with a broad peak 
in efficiency over the ranges $30 < \that < 200$ days and $17 < V < 19.5$.
The function $\epsilon(\that,V)$ falls off rapidly for magnitudes
smaller than $V\sim\,17$
due to an explicit cut $V > 17$ (criteria set B), but the more
gradual fall off for dim stars is a natural consequence of fainter,
lower S/N events.  The gradual drop in efficiency for short duration events
is caused by the sampling issues discussed in \S~\ref{samplingresults}.
The sharp cut at long durations ($\that < 600$ days) seen in the
sampling efficiency has here disappeared, although a remnant can still
be seen for bright events.  The cause of this, of course, is blending.
Bright events are far less likely to be heavily blended, and thus are
unlikely to be recovered with $\that$'s longer than 600 days.
Events that are fainter are more commonly blended (fit $\that$'s much shorter)
and can be scattered below the cut at 600 days.  A corollary to this is 
that intrinsically short events are unlikely to be detected on faint stars,
which are typically blended, as is also seen in Figure~\ref{fig_eff_per_bin},
panel (b).

Given the uncertain knowledge of the source star's luminosity, the somewhat
noisy function $\epsilon(\that,V)$, and the fairly
well constrained shape of the LMC's LF, we have opted to 
integrate out the variable $V$ in the function $\epsilon(\that,V)$.  
Although the shape of the LMC's LF is well known, the overall
normalization to our fields, that is the number of stars with $V < 24$ in our
fields, is less certain (\S~\ref{sec-cmd}).  This uncertainty in 
normalization translates into an uncertainty in the survey's
exposure in \textit{star-years}.  However, it is expected that the efficiency
times the exposure will converge at some magnitude 
$V_{\rm{stop}}$.  Furthermore it is convenient to refer to our `exposure' 
in units of objects-years (the number of lightcurves monitored times the
duration of the survey) as this number is well known.
We opt to move the (uncertain) normalization, that translates the
`exposure' in \textit{object-years} into exposure in \textit{star-years},
into the efficiency.  We can understand this as follows.  Assume a general distribution of event
durations $d\Gamma/d\that$ for some Galactic model.  The number of 
expected events $N_{exp}$ is just,

\begin{equation}
N_{exp} = E_* \int_{0}^{\infty} {\cal E}_* (\that)\,
{d\Gamma \over d\that}\,d\that,
\label{equation12}
\end{equation}

\noindent
where the exposure here, $E_*$, is in star-years and the efficiency is
calculated from $\epsilon(\that,V)$ as,

\begin{equation}
{\cal E}_*(\that) = \int^{V_{\rm{stop}}}\epsilon(\that,V)\,\phi(V)\,dV.
\label{equation13}
\end{equation}

\noindent
Here $V_{\rm{stop}}$ is some cut-off magnitude where the integration is stopped and
$\phi(V)$ is the LMC's LF normalized such that,

\begin{equation}
\int^{V_{\rm{stop}}} \phi(V)\,dV\,=\,1.0. 
\label{equation14}
\end{equation}

\noindent
The exposure, $E_*$, in Equation~\ref{equation12} must be related to the
observed number of objects, $N_{\rm{obj}}$, as,

\begin{equation}
E_* = [S/O(V_{\rm{stop}})] N_{\rm{obj}} T,
\label{equation15}
\end{equation}

\noindent
where $T$ is the time-span of the survey and $S/O(V_{\rm{stop}})$ is a scaling
factor that converts the average number of objects seen in the survey 
to the actual number of stars, down to the cut-off magnitude $V_{\rm{stop}}$.
$S/O(V_{\rm{stop}})$ may be estimated from,

\begin{equation}
S/O(V_{\rm{stop}}) = {\int^{V_{\rm{stop}}} \Phi(V)\,dV \over 
{\int^{V_{\rm{stop}}} \Phi_{\rm{obj}}(V_{\rm{obj}})\,dV_{\rm{obj}}}},
\label{equation16}
\end{equation}

\noindent
where $\Phi_{\rm{obj}}(V_{\rm{obj}})$ is the ground-based luminosity
function and is normalized to the total number of objects observed by the
survey: \ie, $\Phi_{\rm{obj}}(V_{\rm{obj}}) = N_{\rm{obj}}
\phi_{\rm{obj}}(V_{\rm{obj}})$.  Similarly $\Phi(V)$ is the true
underlying LF normalized as in \S~\ref{sec-lfnorm}.
The normalization $S/O(V_{\rm{stop}})$ was estimated for each field in 
\S~\ref{sec-cmd} (Table~\ref{tab-norms} and Figure~\ref{fig-example-norm-lf})
for two cut-off magnitudes $V_{\rm{stop}}\,=\,22$ and 
$V_{\rm{stop}}\,=\,24$ and for two LF ($LF_{1}$ and 
$LF_{2}$).  We make a simple re-definition; let 
${\cal E}(\that) = [S/O(V_{\rm{stop}})] {\cal E}_*(\that)$ and 
$E = N_{\rm{obj}} T$.  With these definitions Equation~\ref{equation12} can 
be written as,

\begin{equation}
N_{exp} = E \int_{0}^{\infty} {\cal E}(\that)\,{d\Gamma \over
d\that}\,d\that. 
\label{equation17}
\end{equation}

\noindent
A similar re-definition may be performed on Equation~\ref{equation2} for the
optical depth.  The advantage of using $E$ and ${\cal E}(\that)$ instead 
of $E_*$ and ${\cal E}_*(\that)$ is twofold.  Firstly, the exposure in
\textit{object-years}, E, is known accurately.  Secondly, this
substitution leaves ${\cal E}(\that)$ containing the only reference 
to $V_{\rm{stop}}$ (implicitly) and we can investigate its 
convergence with magnitude easily.  It is important to note
that with the above definition of ${\cal E}(\that)$, the
efficiency is no longer bound to lie below one.  This slightly 
un-intuitive result
is due to the fact that $S/O(V_{\rm{stop}})$ may be quite large 
(Table~\ref{tab-norms}), though in practice ${\cal E}(\that)$
always lies below one.  A practical way of viewing ${\cal E}(\that)$ is:
given $S/O(V_{\rm{stop}})$ events, ${\cal E}(\that)$ is the expected
number of detected events for the given $\that$.  In the limit 
$S/O(V_{\rm{stop}}) \rightarrow 1.0$ the efficiency ${\cal E}(\that)$ 
recovers its usual meaning.

Figure~\ref{fig_eff_per_bin} illustrates the integration of
$\epsilon(\that,V)$, panel (b), over a LF, panel
(a).  The resulting function ${\cal E}(\that) =
[S/O(V<24)]*{\cal E}_*(\that)$ is shown in panel (c).  Our two
LFs, $LF_{1}$ and $LF_{2}$, and their corresponding
values of $S/O(V<24)$ are displayed as dotted and solid lines, 
respectively.
An important result is that ${\cal E}(\that)$ is fairly robust to
uncertainties in the LF fainter than $V\sim\!22$.
For durations less than 75 days the difference in ${\cal E}(\that)$
as derived using either LF is less than 1\%.  The
difference, however, becomes progressively larger for longer durations
due to the relative difference in contributions from faint stars exhibited
by the two LFs.  At 300 days the difference
is $\sim\!3\%$.  In \S~\ref{sec-cmd} we chose to favor $LF_{2}$ over 
$LF_{1}$ because of our HST completeness tests and because it
seems unlikely that the LF will continue to rise so steeply beyond
the clump for so long.  Evidently any moderately different LF would
produce only a small difference in the overall efficiency. 

In Figure~\ref{fig_phot_eff} we present the efficiency
${\cal E}(\that)$ for selection criteria set A and B using LF
$LF_{2}$.  Also shown for comparison are the photometric efficiencies used in 
\yrone\ and \yrtwo.  Many of the differences between the old
and new results, as seen in the sampling efficiencies 
(\S~\ref{samplingresults}),
can also be seen here.  Again the most striking difference is the much
larger efficiency at long durations.  As described in \S~\ref{samplingresults}
this has multiple causes, including (1) almost three times more baseline in
the \yrfive\ data than in the \yrtwo\ data, (2) a longer duration cut of
$\that < 600$ days and (3) 6 high density fields which were previously split 
into two separate years of data (due to an early generation of templates)
have been recombined.  A new reason, unique to the photometric efficiency,
is the contribution of faint stars to the efficiency for long duration
events.  This effect was troublesome in the 
\yrone\ and \yrtwo\ results since we lacked adequate numbers of faint
stars (only a few with $V\sim\!22$ and none with $V > 22$) and thus were 
only confident that the photometric efficiency had converged for durations
less than 150 days.  As we show below the \yrtwo\ results were likely to have
converged only for durations less than 100 days.  However, this resulted in 
an underestimate in the efficiency of less than 10\% for
durations around 150 days.  Criteria set A is closer in design to the
\yrtwo\ cuts and, as a consequence, follows the older results more
closely for very short durations up to about 60 days. 
As noted above the sharp cut-off at durations around 600 days seen in the
sampling efficiency is smoothed over in the photometric efficiency 
because blending scatters intrinsically long duration events to
shorter measured durations.

\subsection{Convergence}
\label{sec-convergence}

Figure~\ref{fig_convergence} shows the convergence of ${\cal E}(\that)$
with $V_{\rm{stop}}$ for the four combinations of selection 
criteria sets A and B and $LF_{1}$ and $LF_{2}$.
In each panel the convergence of four different durations  
($\that = $ 50, 100, 300, and 1000 days) is shown.  The ordinate is 
in arbitrary relative units.
Criteria set B is our loosest set of cuts and is the least convergent
of the two selection criteria sets used in \yrfive.  An inspection of
the figure also shows that $LF_{1}$ gives a somewhat slower
convergence than $LF_{2}$.  This is not surprising as $LF_{1}$
contributes a substantially larger number of fainter stars as compared
with $LF_{2}$.  Criteria set A with $LF_{2}$ converges for durations less 
than 1000 days by
$V_{\rm{stop}}\sim\!24.5$ and criteria set B with $LF_{2}$ converges
for durations less than 300 days by $V_{\rm{stop}}\sim\!24$.  
If the true LF is closer in form to 
$LF_{1}$ then it is possible that our efficiency results are
somewhat underestimated for long duration events (see 
\S~\ref{sec-errorbudget} below).  However, since none of the 17 candidates 
in \yrfive\ have durations longer than 300 days we feel that the 
convergence of the efficiency is more than adequate.

\subsection{Error Budget}
\label{sec-errorbudget}

There are a number of potential sources of error in estimating
our efficiency and we list the most important
ones in Table~\ref{tab-errorbudget}.  We crudely classify the errors
as `signed' or `unsigned'.  That is, if we are reasonably certain
the effect would only increase (or decrease) the efficiency we classified 
it as `signed' and marked it with the appropriate sign in the table.
If we are not certain of it's sign we called it `unsigned' and left
a question mark in the sign column of the table.

The first unsigned error in the table (\#1) is simply the uncertainty in
our normalizations and was estimated to be $\sim\!20$\% in 
\S~\ref{sec-lfnorm}.  That is we are confident that we know the 
number of stars to the limit $V=24$ in our LMC fields to within 
$\sim\!20$\%.  This is by far our largest source of error in the
efficiency.  The second unsigned error (\#2) is an uncertainly 
in our incomplete knowledge of the shape of the LMC's LF
at faint magnitudes.  In \S~\ref{subsec-phoeff} we used the 
difference between $LF_{1}$ and $LF_{2}$ to estimate a 1-3\% effect 
depending on the duration of the event.  Another source of unsigned
error (\#3) is due to the finite number of Monte Carlo events, which 
we estimate to be $\simlt\,1$\% based on the binomial distribution and
the number of Monte Carlo events used.

The fourth entry in Table~\ref{tab-errorbudget} (\#4) is signed and
represents a potential over-completeness bias.  Because blending
might over-populate objects with respect to stars in our calibration 
bin ($17.5 < V < 18.5$), this could cause an overestimation of
${\cal E}(\that)$.  In \S~\ref{sec-lfnorm} we estimated
this could be as large as $\sim\!9$\%, but likely to be somewhat
smaller (perhaps 6-8\%).  The next signed error (\#5) is related 
to our choice of using only the nearest neighbor to each artificial
star, when in fact as many as $\sim\,3$\% of PRFs had 2nd or 3rd
nearest neighbors which were more sensitive to the lensed flux.
Of course what matters is how many 2nd or 3rd nearest neighbors would
have been detected but were not because the closest neighbor was used
instead, and this is probably smaller than the number of 2nd and 3rd nearest 
neighbors who are more sensitive.  The efficiency is likely to be 
underestimated somewhat by this error.  The next two signed 
errors are small and relate to (\#6) our use of a `universal' LF 
(especially in the outer LMC disk fields where the true LF likely
turns over at brighter magnitudes) and (\#7) a possible slight
underestimation of ${\cal E}(\that)$ if convergence has not been
reached by $V = 24$ (\eg, if the the true underlying luminosity
function is closer to $LF_{1}$; \S~\ref{sec-convergence}).

We leave two potential sources of error unexplored in this paper
because of their highly uncertain and complicated nature.  The
first of these (\#8 in the table) is the difficulty of incorporating
binary star systems into the determination of the LF.  Binary stars
could increase our exposure (thereby increasing ${\cal E}(\that)$)
but also shift the LF toward fainter magnitudes (decreasing 
${\cal E}(\that)$).  The effect on the LF of the local neighborhood
is still controversial (\cite{kroupa,reid}) and any hope
of resolving
this for the LMC is beyond the scope of this paper, and left 
for future work.  Another unexplored source of error (\#9) is due to
our lack of exotic lensing Monte Carlo events.  The behavior of this 
source of error is difficult to estimate as the shape of exotic lensing
is likely to lower our efficiency, but caustic crossings might well 
increase it (\cite{distefano1,distefano2}).  Binary lensing events are 
likely to dominate this effect, but the distributions of their event
parameters is wide open to speculation (but see Alcock \etal 1999c).
We plan to investigate the effects of a binary 
lens population on our efficiency in a future paper.

Given the generally small size of the individual signed errors
relative to the size of the unsigned errors, as well as
the Poisson-like `counting' errors in the optical depth
estimates (13-17 events gives $\sim\,30-40\%$ errors in the optical depth;
see Table~10 in \yrfive), halo mass fraction, etc.,
and the fact that a number of these signed
errors are of comparable size but opposite sign, we have chosen not
to attempt to correct for them.  Our estimates of their size
are rough, but probably accurate to within a factor of two.
There is no correct way to total systematic errors - but assuming 
they are uncorrelated we can sum them in quadrature to get a total 
of $\sim\!22$\%, not including errors \#8 and \#9, which may be the largest,
but for which we do not have good estimates.  Alternatively, we can 
add the signed errors algebraically to get $\sim\!5$\%, then to this add 
the unsigned errors in quadrature, giving $\sim\!21$\% total error.
So our best guess at the size of our error is 21-22\% plus the unknown 
effects of binary corrections to the LF and exotic microlensing.

\subsection{Efficiency Differences Across Fields}
\label{sec-perfield}

Our efficiency for each field is shown in 
Figure~\ref{fig_photeff_vs_field}.  These have been calculated in the
manner described above and normalized with each field's $S/O$ 
(Table~\ref{tab-norms}).  By splitting 
our $\sim\!1.2$ million Monte Carlo events into 30 fields, the 
shot noise has increased to 5-10\%, which is somewhat field dependent.
In principle the shot noise can be reduced by simply running more passes,
but the CPU time required is somewhat prohibitive at the present time.
Nevertheless the efficiency for each field is important when looking for
gradients in optical depth or rates across the face of the LMC, which
the present level of accuracy is adequate for, given the small number of
events (13-17) seen toward the LMC.  Differences seen
between fields in Figure~\ref{fig_photeff_vs_field} are due primarily to
(1) different normalizations for each field (Table~\ref{tab-norms}),
physically corresponding to larger numbers of stars and thus probability
of detecting events in some fields, (2) different sampling rates due to 
observing strategy and weather across fields, and (3) to a smaller 
degree more crowded fields tend to have a larger effective area (area covered
by stellar PSFs in the image plane) in which to detect the presence of
magnified events, because of blending.
 
\subsection{Parameter Bias}
\label{subsec-parameter-bias}

As discussed in the introduction, blending induces a bias in the 
measured parameters $\Amax$ and $\that$.  We can quantify this
bias using our Monte Carlo events, and have developed a method of
statistically correcting for $\that$ bias when calculating optical
depths.  Figure~\ref{fig_bias} plots the ratio 
$\rAmax/\Amax$ versus $\rthat/\that$ for a random 
sample ($N\,\sim\,10^{5}$) of Monte Carlo events: panel (a) is for sampling
efficiency events and (b) is for photometric efficiency events (both use
selection criteria set B).  The majority of sampling efficiency
events are recovered with little or no parameter bias in $\that$.
(There is a small amount of bias in $\Amax$ (both directions) caused
by sampling and weather gaps that effectively obfuscate very high
magnification events.)  As the sampling efficiency does not simulate
blending this result is not surprising.  The results for the photometric
efficiency events is quite different.  A significant amount of blending
can be seen in panel (b) of Figure~\ref{fig_bias}, were approximately
40\% of the Monte Carlo events are blended with an additional 10\%
or more of un-lensed flux.  The trend of decreasing $\Amax$ with decreasing
$\that$ for highly blended events is well delineated in the figure.
Note that the photometric efficiency events sample the CMD (luminosity
function) in a square-root fashion (\S~\ref{sec-prdb}) and thus dim 
stars are underpopulated in panel (b) of Figure~\ref{fig_bias}.  Factoring
this in, as is done below, further increases the amount of blending
and the size of any potential correction.

We can calculate the efficiency as a function of various parameters
(\eg, $\rumin$, $\umin$, $V$, $V_{\rm{obj}}$, etc.) for the
photometric efficiency, and some of these are shown in \yrfive\
(for example, see their Figure~6).  However, since the Monte Carlo 
events sample the CMD in a square-root fashion we must bin and 
re-weight with the correct LF.  We must also bin 
and re-weight with a realistic distribution in $\that$ as the 
Monte Carlo events are added with durations uniform in $log(\that)$,
which is unlikely to be the true distribution.  We chose to re-weight
with the distribution of durations predicted by a standard halo model 
with delta function mass $0.5\,M_{\odot}$ (\cite{griest91}), since 
this distribution closely matches the data (average duration
$\that\!\sim\!92$ days; see \yrfive).  Fortunately, model dependency is 
weak as the distributions in $\umin$, weighted and un-weighted
(that is using the $log(\that)$ distribution) are quite similar. 

If the $\that$ bias is left uncorrected the optical depth will be
underestimated (Equation~\ref{equation2}).  There are
two very different, but complimentary, ways to correct for this bias in 
$\tau_{meas}$.  Ideally one could perform microlensing fits on individual events
that allow for unlensed flux in each passband.  Then each
event could be corrected for blending separately and the maximum amount 
of information in the set of $\that$'s would be retained.  In practice
this approach is difficult as the lightcurves are not always well sampled
and can have considerable photometric scatter making the blend fits 
uncertain (\cite{lmc1,pratt,wozniak,han}).  This situation can be
greatly improved with accurate, dense, follow-up photometry on alerted
events as is being done now routinely (\cite{albrow-planet,95-30,rhie-mps}).
However, not all events are alerted.
In addition, blending fits
are not always unique and only on fairly high magnification events can
the fit parameters be extracted with confidence.
A second approach is to correct the
$\that$ bias in a statistical fashion.  Since we have \textit{a priori} 
knowledge of the distribution of source stars from the luminosity 
function $\phi(V)$ and can estimate the distribution of uncorrected
durations ${d\Gamma/d\that}(\that)$ we can compute a first order
average correction as,

\begin{equation}
\overline{\alpha} = {{\int\,dV\,\int\,d\that\,\,\alpha(\that,V)\,\epsilon(\that,V)\,
{d\Gamma \over d\that}(\that)\,\phi(V)}\over{
{\int\,dV\,\int\,d\that\,\,\epsilon(\that,V)\,
{d\Gamma \over d\that}(\that)\,\phi(V)}}},
\label{equation18}
\end{equation}

\noindent
where $\epsilon(\that,V)$ is defined as above
and $\alpha(\that,V)$ is defined as: $\alpha(\that,V) =
median(\rthat/\that)$ or the median $\that$ bias for the
Monte Carlo events in a bin of $\that$ and $V$.
The quantity $\alpha(\that,V)$ lies between zero and one and is
a strong function 
of $\that$ and $V$ in the sense that $\alpha(\that,V)$ is one for bright 
events and approaches zero for both dim events and for long duration
events.  The results of Equation~\ref{equation18} for criteria set A and B
using LFs $LF_{1}$ and $LF_{2}$ are tabulated in 
Table~\ref{tab-bias}, assuming mean durations of 41, 92, and 130 days 
(corresponding to delta-function masses of 0.1, 0.5, and 1.0 $M_{\odot}$
in the model ${d\Gamma/d\that}(\that)$,
respectively).  The mean duration of the microlensing candidates
in \yrfive\ corresponds most closely with our value of $\that\,\sim\,92$
days or an average mass of $M_{\odot}\,\sim\,0.5$.  This correction
may be used in Equation~\ref{equation2} by simply substituting 
$\rthat\,\rightarrow\,\rthat/\overline{\alpha}$ to obtain,

\begin{equation}
\tau_{meas}\rightarrow\,{{\pi}\over{4}}{{1}\over{E}}\sum_i{{\that_i'}/
\overline{\alpha} \over{{\cal E}(\that_i'/\overline{\alpha})}}.
\label{equation19}
\end{equation}

\noindent
Note the optical depth defined in this manner does
not scale simply as $1/\overline{\alpha}$.  We have tried computing
$\overline{\alpha}$
as a function of measured parameters (such as $V_{\rm{obj}}$ 
and $\rthat$)
but find that it is fairly constant over the measured ranges.
The use of a model distribution in $\that$ is somewhat worrisome.  However,
the range in corrections $\overline{\alpha}$ for different $\that$ 
distributions is acceptably small, about 5\% (Table~\ref{tab-bias}).
The error induced in the optical depth is somewhat smaller
($\sim\!3\%$; see Table~9 in \yrfive).

To check this statistical correction we ran a set
of Monte Carlos on a toy model using our artificial events.  The 
Monte Carlos work as follows: (1) first we bin the recovered artificial
events in $\that$ and $V$, (2) we create $N_{exp}$ `events' by sampling 
the LF $\phi(V)$ and $\that$ distribution function 
${d\Gamma/d\that}$ for a simple halo model with a delta function
mass $0.5\,M_{\odot}$ (model S in \yrone, \yrtwo, and \yrfive), (3) for
each of these $N_{exp}$ `events' we chose an artificial event to represent it
by selecting at random from the appropriate bin in $\that$ and $V$, (4)
three different optical depths for these $N_{exp}$ `events' are computed using
three different $\that$'s (the true $\that$, the fit 
$\rthat$ and the statistically corrected $\that_{st}\,=\,\rthat/
\overline{\alpha}$).  The number of $N_{exp}$ `events' corresponds
to the predicted number of observed events as computed by 
Equation~\ref{equation3}.
The experiment is repeated 1000 times and the results for selection
criteria set B and $LF_{2}$ are displayed as histograms
of $\tau_{meas}$ in Figure~\ref{fig_monte_tau}.  The \textit{dashed histogram}
corresponds to calculating $\tau_{meas}$ using the fit $\rthat$'s 
and clearly underestimates the optical depth 
($\tau_{model}\,=\,4.7\times10^{-7}$ for Model S; the vertical line in
the figure).  Using the true
durations $\that$ to compute $\tau_{meas}$ gives the correct optical 
depth (\textit{solid histogram}). 
The optical depth computed using the statistically
corrected durations, $\rthat/\overline{\alpha}$, is displayed as the
\textit{dotted histogram} and correctly predicts the true optical depth
in a fairly unbiased manner.  Similar unbiased results are found using 
selection criteria set A, $LF_{1}$, $LF_{2}$, and 
using the ${d\Gamma/d\that}$ distributions with delta-function masses
of 0.1 and 1.0 $M_{\odot}$.
These Monte Carlos give us confidence that this statistical correction
can be used in an unbiased manner to compute the optical depth toward 
the LMC.

\section{SUMMARY}
\label{sec-summary}

The results of \yrfive\ for 5.7 years of photometric data toward
the LMC rely critically on how well we understand the detection
efficiency.  In this paper we calculated these efficiencies, 
correcting the most important systematics effects with realistic 
models.  To account for the wide range of issues due to blending
we perform artificial star tests on a broad range of images.
These artificial stars tests allowed us to empirically account for 
blending and to model our photometry code's systematics.  Correcting
for blending also required an accurate knowledge of the LMC's 
LF because microlensing magnifies flux and
increases the survey's sensitivity to dim, unresolved stars.

We found that our sensitivity ${\cal E}(\that)$ to dim,
unresolved stars vanishes for magnitudes fainter than $V\sim\!24$
and durations less than $\that\sim\!300$ days.  Our sensitivity 
in previous results (\yrone\ and \yrtwo) was somewhat underestimated
for durations greater than $\that\sim\!100$ days.  We also found the
$\that$ bias, due to blending, is of the order of 20\% and we presented
a method for statistically correcting this bias in our optical 
depth estimates.  The method is complimentary to using blended 
$\that$ fits.

\section*{ACKNOWLEDGEMENTS}

We are very grateful for the skilled support given our project
by the technical staffs at the Mt.~Stromlo and CTIO Observatories,
and in particular we would like to thank S. Chan,  G. Thorpe,
S. Sabine, J. Smillie, and M. McDonald, for their invaluable assistance in
obtaining the data.  We especially thank J.D. Reynolds for valuable
assistance with the database software that has made this effort
possible.

This work was performed under the auspices of the U.S. Department of
Energy by University of California Lawrence Livermore National
Laboratory under contract No. W-7405-Eng-48.
Work performed by the Center for Particle Astrophysics personnel
is supported in part by the Office of Science and Technology Centers of
NSF under cooperative agreement AST-8809616.
Work performed at MSSSO is supported by the Bilateral Science
and Technology Program of the Australian Department of Industry, Technology
and Regional Development.
DM is also supported by Fondecyt 1990440.  CWS thanks
the Packard Foundation for their generous support.  WJS is supported
by a PPARC Advanced Fellowship.
KG is supported in part by the DOE under grant DEF03-90-ER 40546.
TV was supported in part by an IGPP grant.

\clearpage
\onecolumn

\plotone{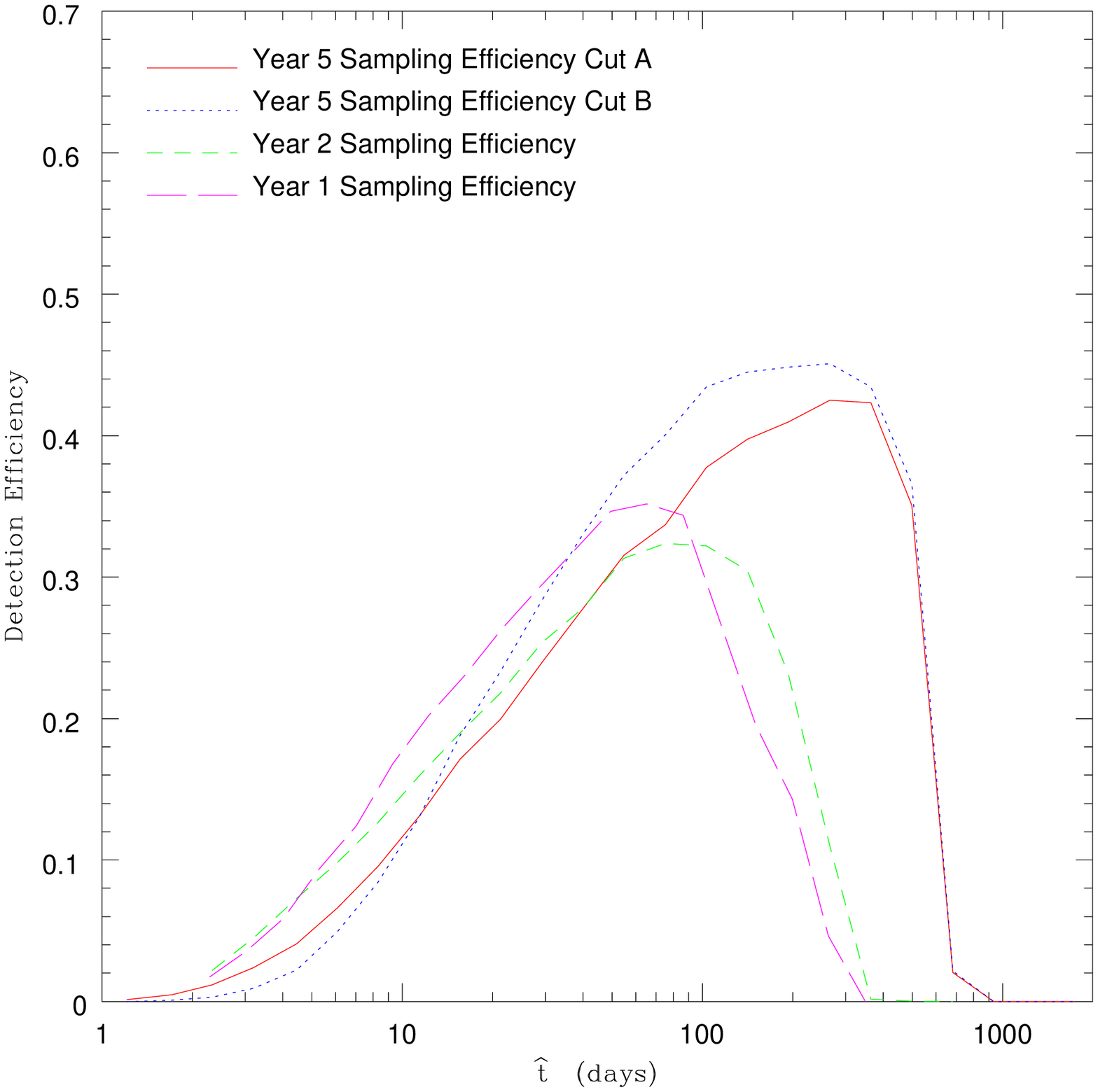}
\figcaption[fig-samp-eff.ps]{
The 5.7-year sampling efficiency (normalized to $\umin < 1.0$)
results for criteria set A (\textit{solid line}) and criteria set B 
(\textit{dotted line}).
Also shown are the sampling efficiencies from \yrone\ 
(\textit{long dash}) and \yrtwo\ (\textit{short dash}) for
comparison.
\label{fig-samp-eff}}

\plotone{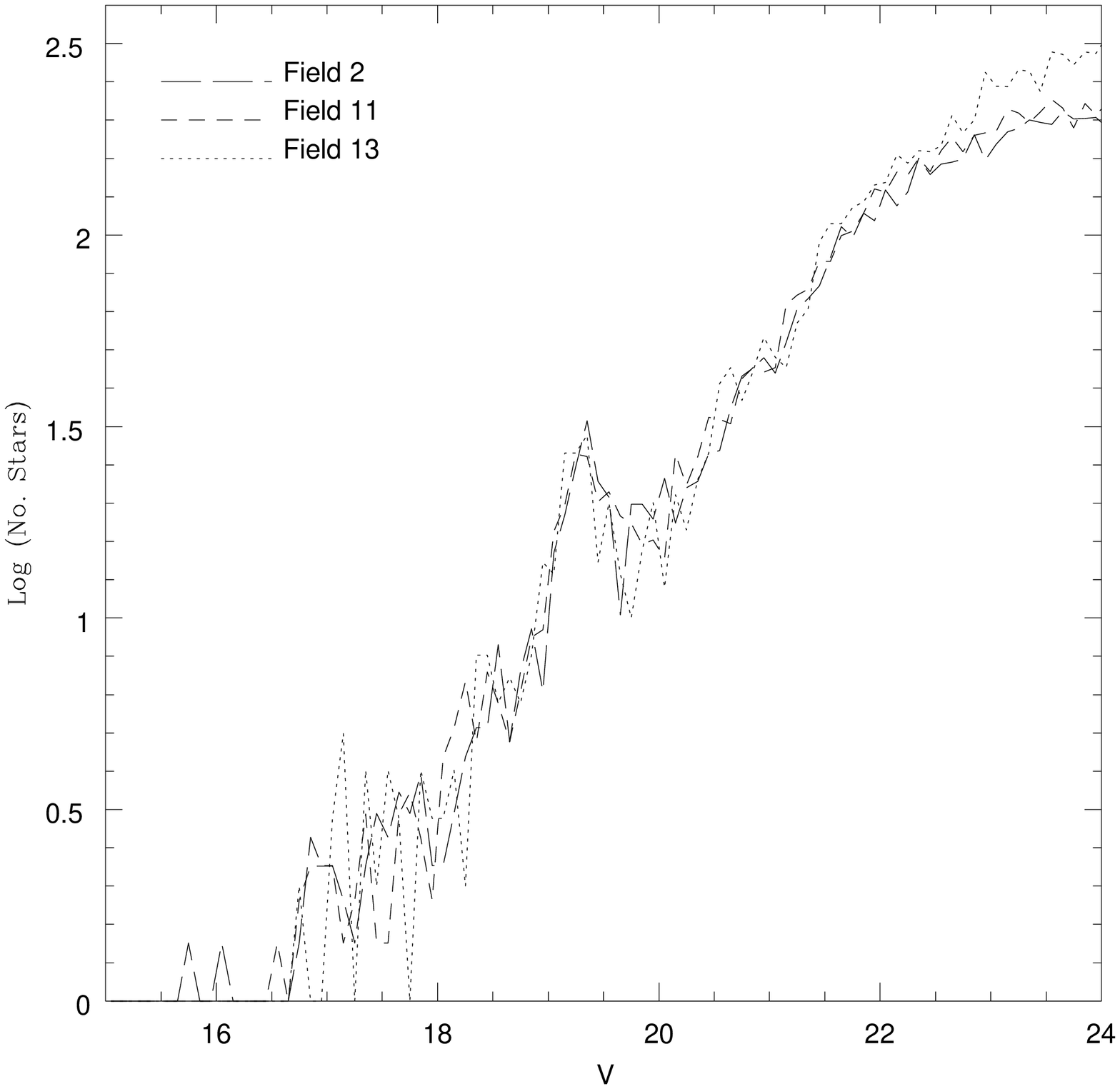}
\figcaption[fig-macho-hst-lf.ps]{
Three HST LFs of MACHO fields are plotted with line codings displayed
in the figure.  Details of the relative normalizations can be found in the
text.  Note the good match in shape from $V\!\sim\!17$ to 
$V\!\sim\!22$.
\label{fig-macho-hst-lf}}

\plotone{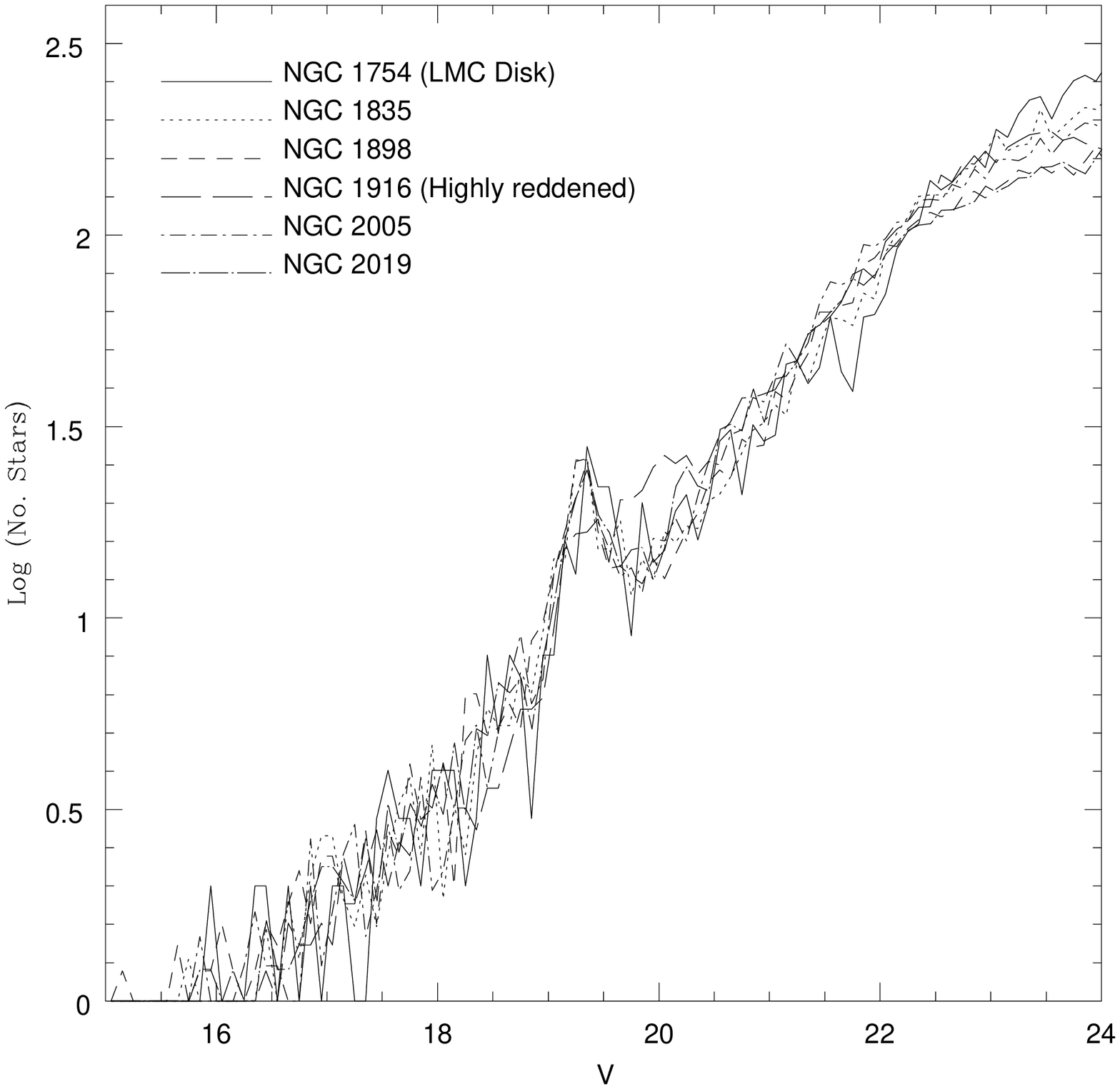}
\figcaption[fig-olsen-hst-lf.ps]{
Six Olsen (1999) HST LFs are plotted with line codings as
displayed in the figure.  Details of the relative normalizations
can be found in the text.  Note the good match in
shape from $V\!\sim\!17$ to $V\!\sim\!22$.
\label{fig-olsen-hst-lf}}

\plotone{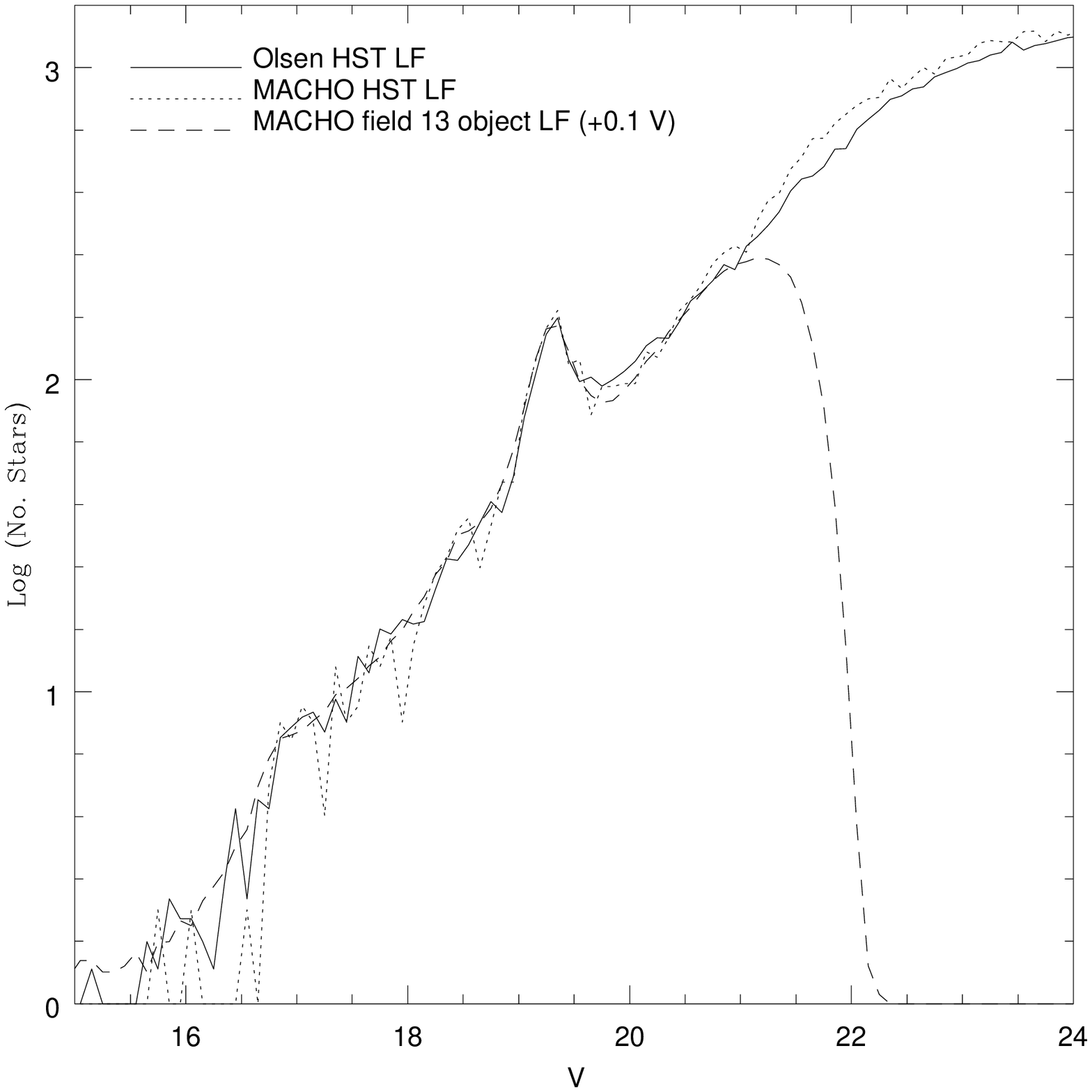}
\figcaption[fig-olsen-macho-f13-lf.ps]{
Combined MACHO HST LF (\textit{dotted line}) is plotted
and compared with the combined Olsen (1999) HST LF 
(\textit{solid line}).  Also shown is the ground-based LF 
for MACHO field 13 (\textit{dashed line}) which have been
normalized as described in the text.  The agreement in
shape over the range $17 < V < 21$ is good.
\label{fig-olsen-macho-f13-lf}}

\plotone{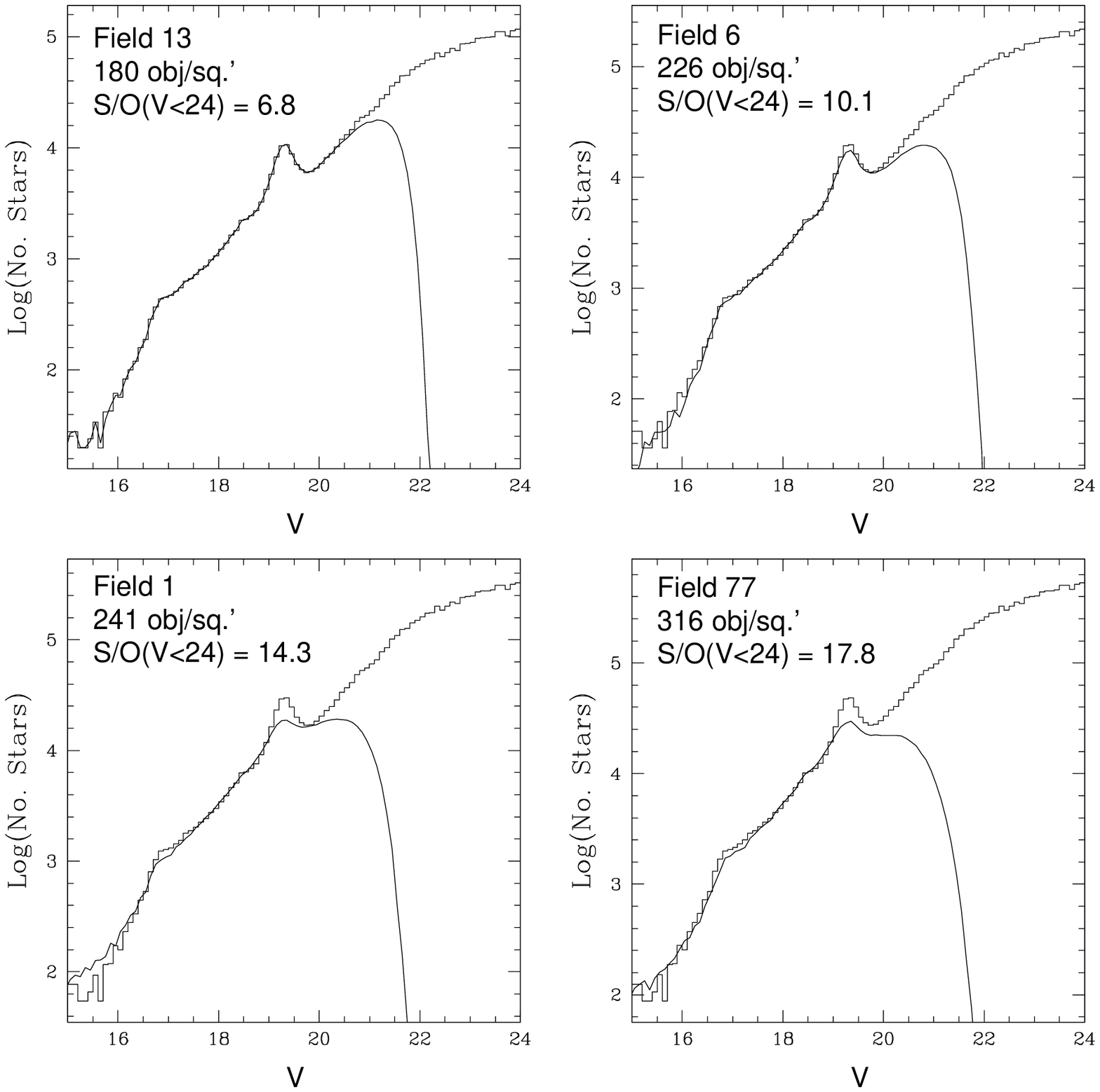}
\figcaption[fig-example-norm-lf.ps]{
Four sample MACHO ground-based LFs (\textit{solid lines}) ranging from low
to high stellar density are labeled in the figure.  $LF_{2}$
(\textit{histograms}) has been offset and normalized
(Table~\ref{tab-norms}) to each field.  The agreement in shapes between 
the individual ground-based LFs and $LF_{2}$ is quite 
reasonable.
\label{fig-example-norm-lf}}

\plotone{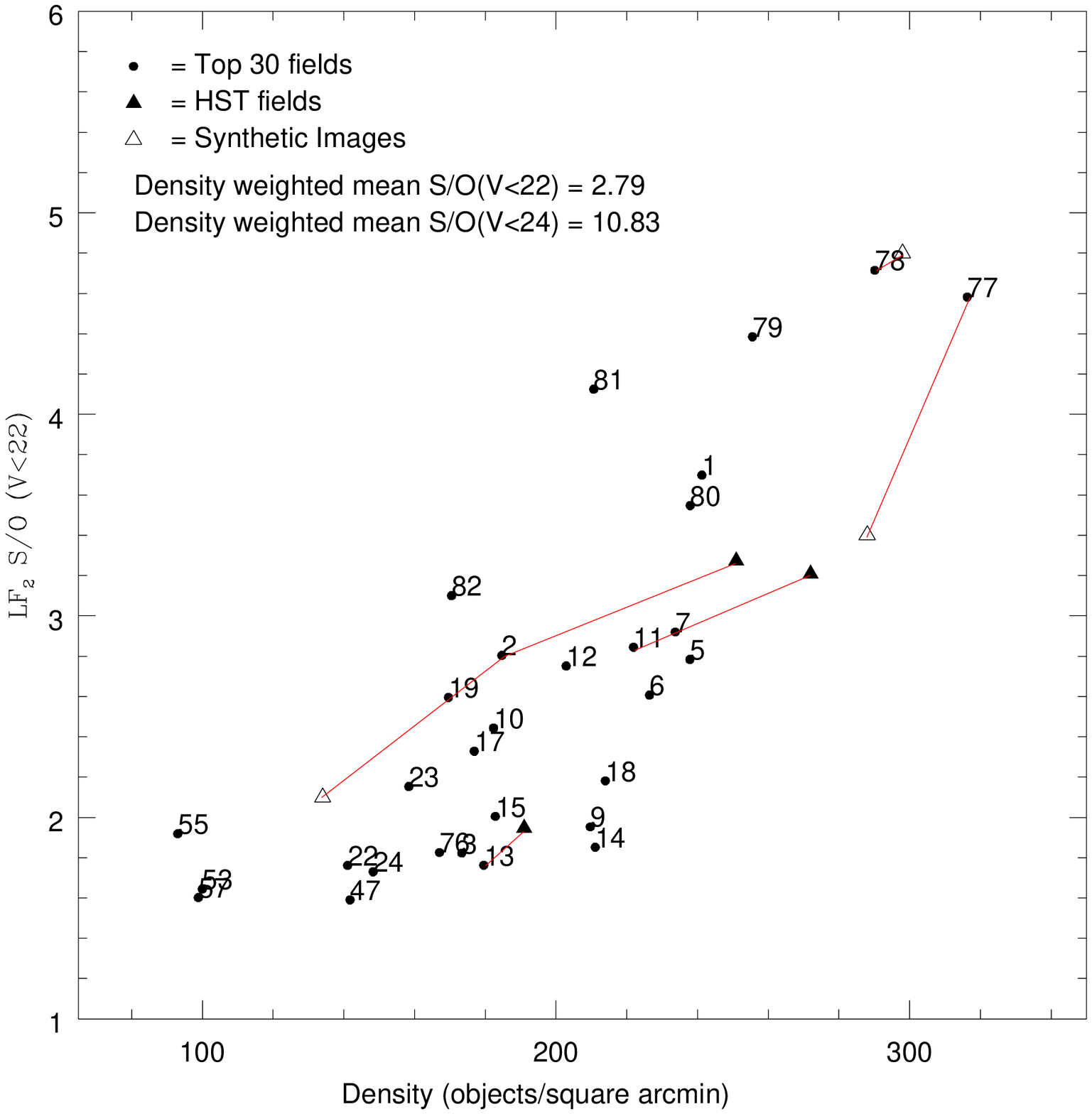}
\figcaption[fig-so22-density.ps]{
Normalization of each field as a function of object density. See
text for details.
\label{fig-so22-density}}

\plotone{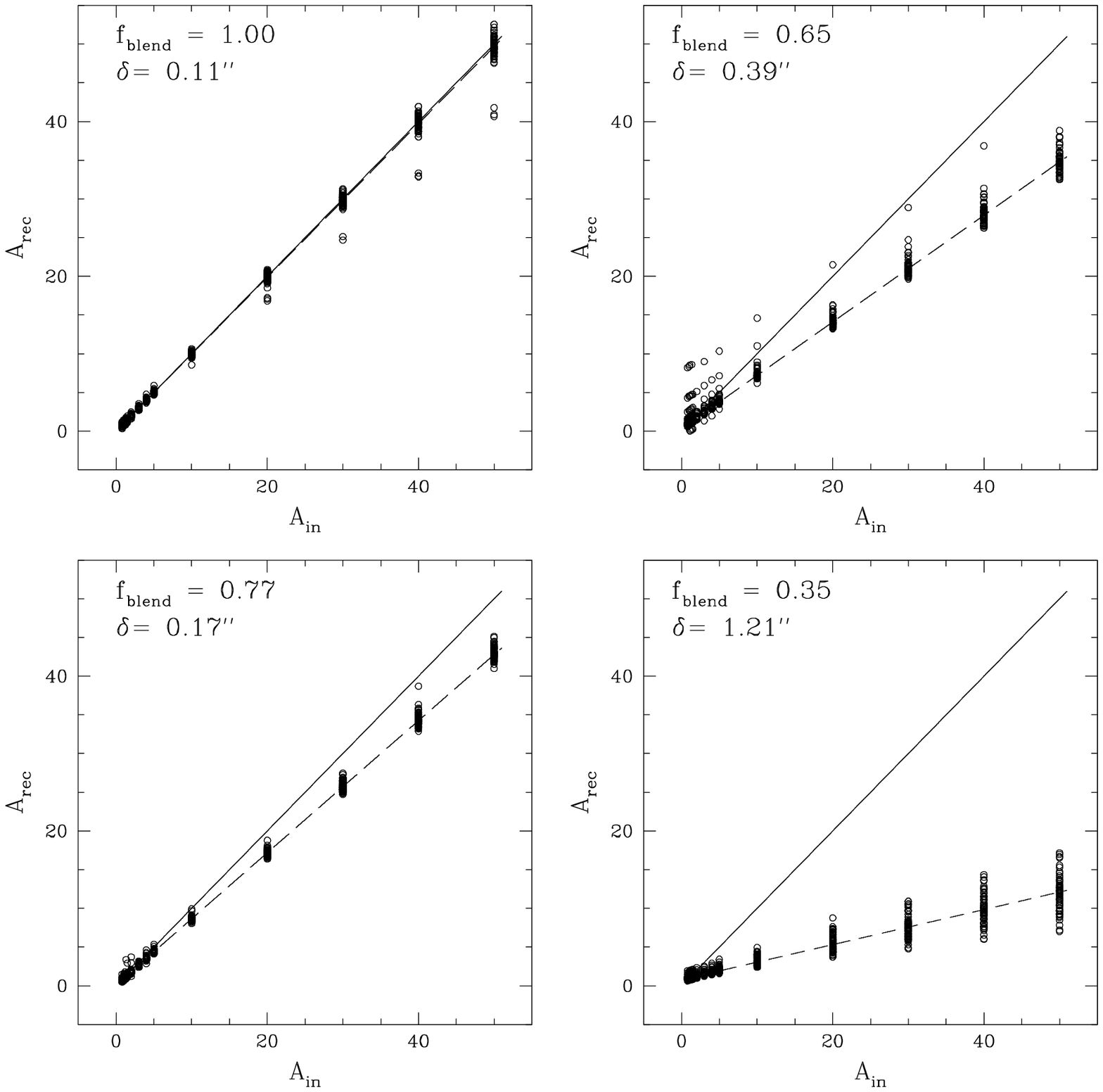}
\figcaption[fig-prf-magnif_77.ps]{
Four photometric response functions (PRFs) are plotted as recovered
magnification $A'$ versus input magnification $A$
(\textit{open circles}) and compared with the ``sampling efficiency"
analytic response function described in \S~\ref{sec-sampling}
(\textit{solid lines}).  The \textit{dashed line} is a least squares
fit to the PRF data and in each panel the fit slope (fraction of `lensed
light' or blending fraction $f_{blend}$) is labeled along with the 
distance ($\delta$ in arcsecs) of the `lensed' star to the nearest
photometred object.
\label{fig-prfs}}

\plotone{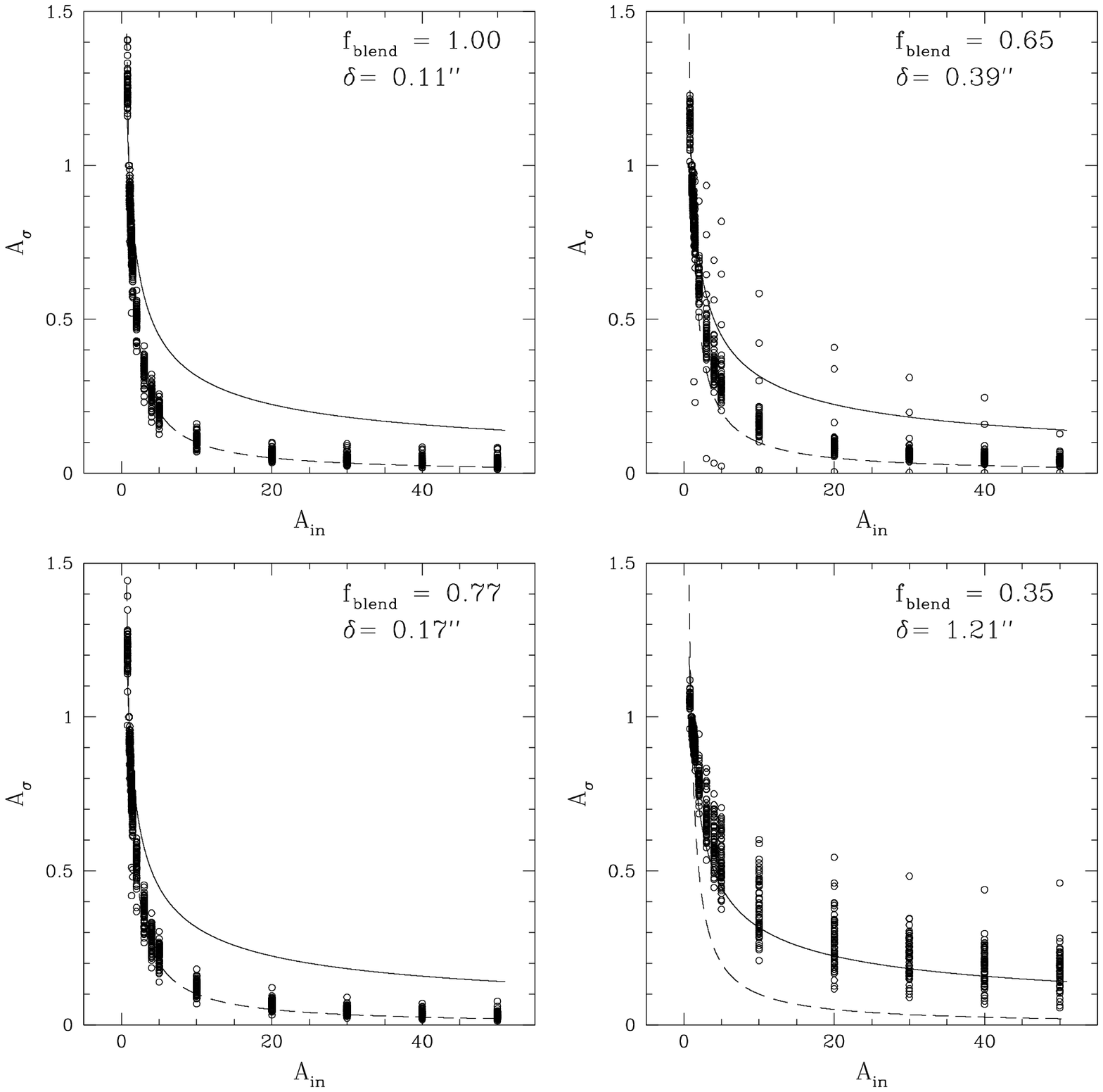}
\figcaption[fig-prf-error_77.ps]{
The four panels here correspond to the same PRFs shown in
Figure~\ref{fig-prfs}.  The characteristic change 
in the relative error (error in magnitudes) $A_{\sigma}$ is
plotted versus $A$.  Overplotted on each panel is
the purely Poisson response assumed in the sampling efficiency
(\textit{solid line}; see \S~\ref{sec-sampling}) and the 
behavior $\sim\!1/A$ (\textit{dashed line}).  See the 
text for more explanation.
\label{fig-prferrors}}

\plotone{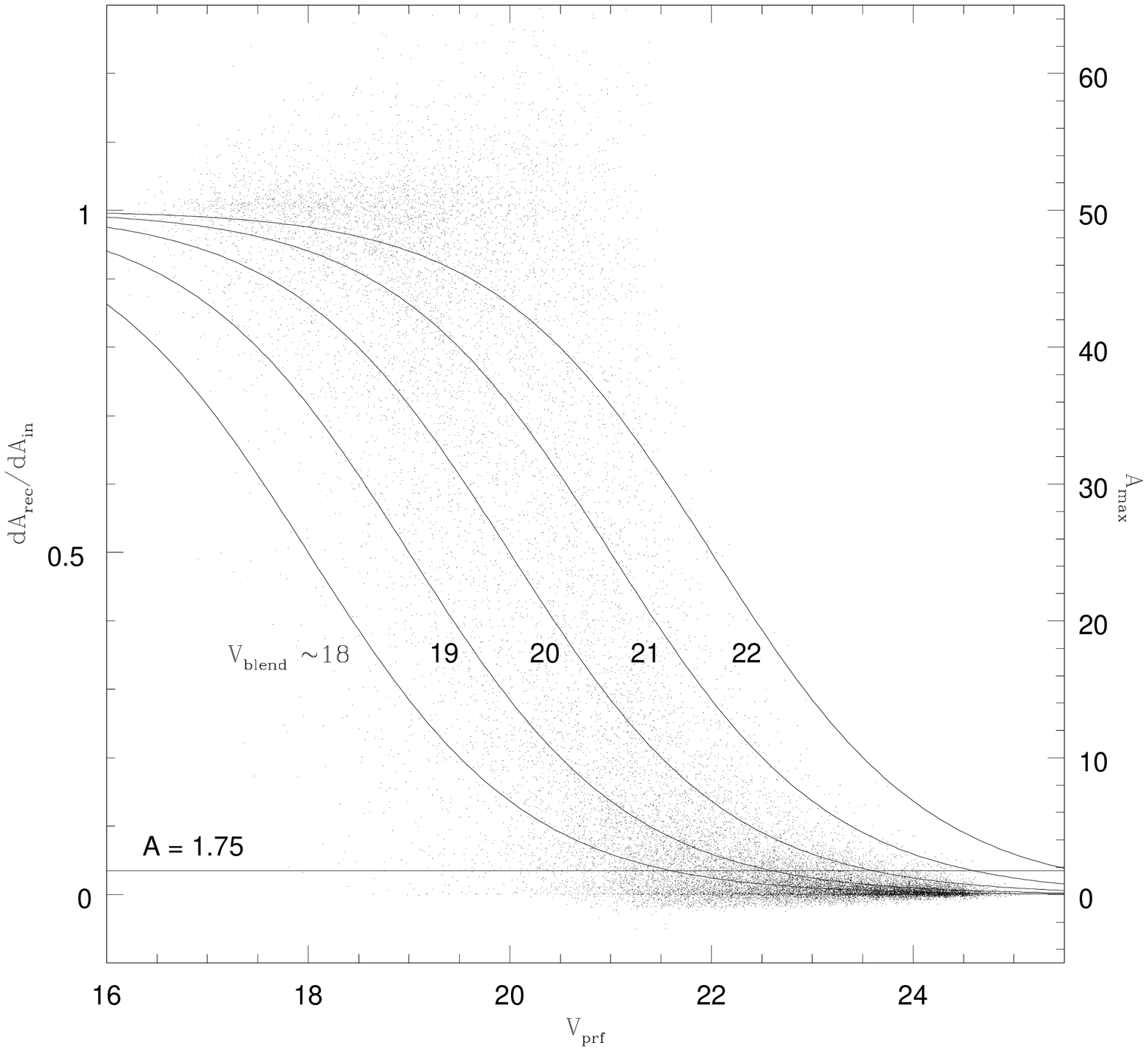}
\figcaption[fig_slope.ps]{
Plot of effective blend fraction (fit $m = dA_{rec}/dA$)
versus $V$ magnitude for the $\sim\!59,000$ PRFs contained
in the PRDB.  Blending is not limited to faint stars, but clearly
becomes worse for fainter magnitudes.  Note that the 
distribution of PRFs shown 
here samples the LF in a square-root fashion 
(for reasons discussed in the text) and as such dim stars shown here
are under-weighted.  The family of curves and the right--hand axis are
explained in the text.
\label{fig-slopes}}

\plotone{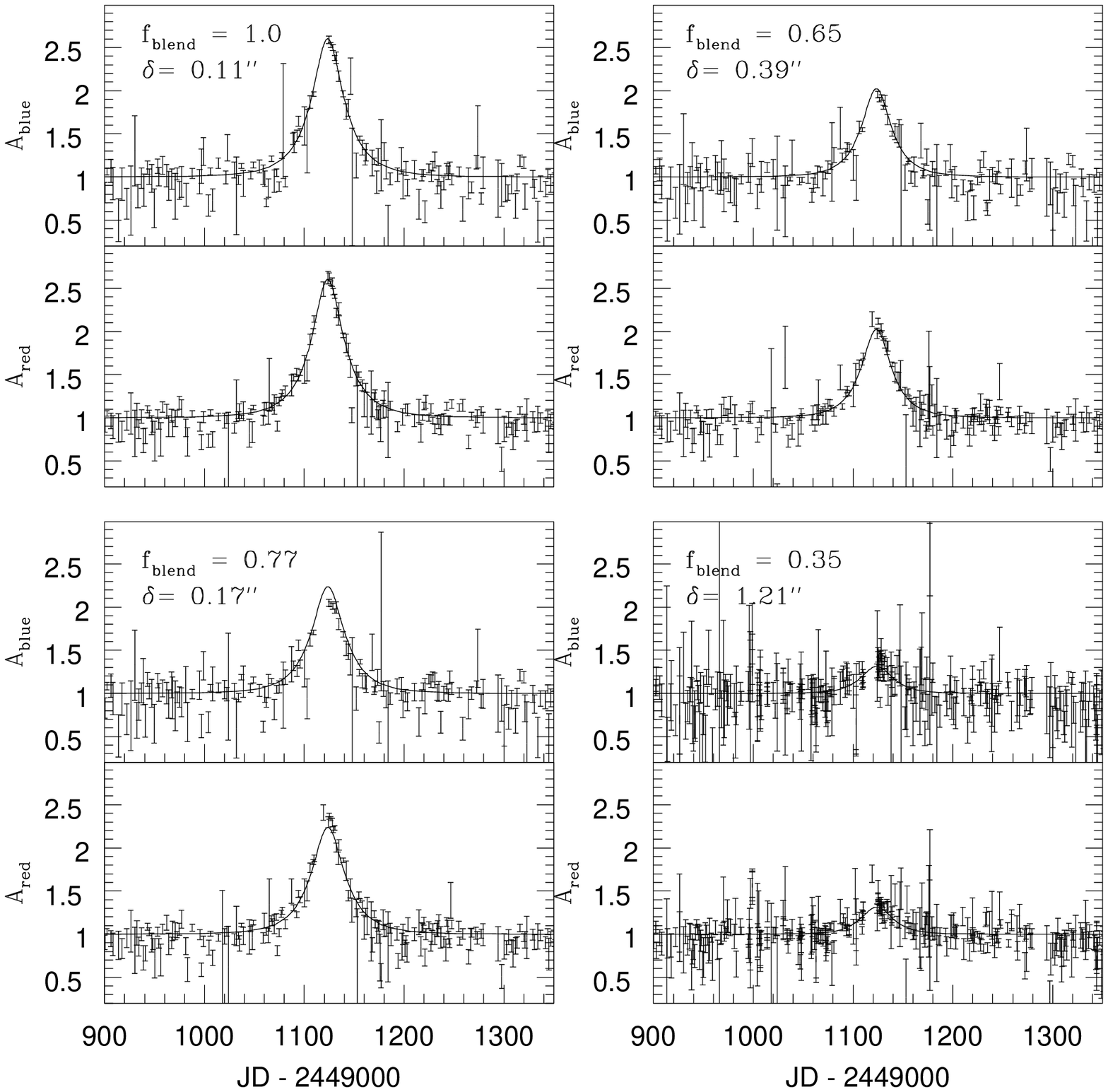}
\figcaption[fig-artif-lc_77.ps]{
Displayed here are four example Monte Carlo lightcurves with identical
input event parameters ($\Amax = 2.2$ and $\that = 80$ days) generated
using the same four PRFs displayed in Figures~\ref{fig-prfs}~and~\ref{fig-prferrors}. 
The \textit{solid line} is the standard microlensing fit described in
\S~\ref{sec-sodo}.  The error bars are $\pm\,1\sigma$.
\label{fig-artif-lc}}

\clearpage

\plotone{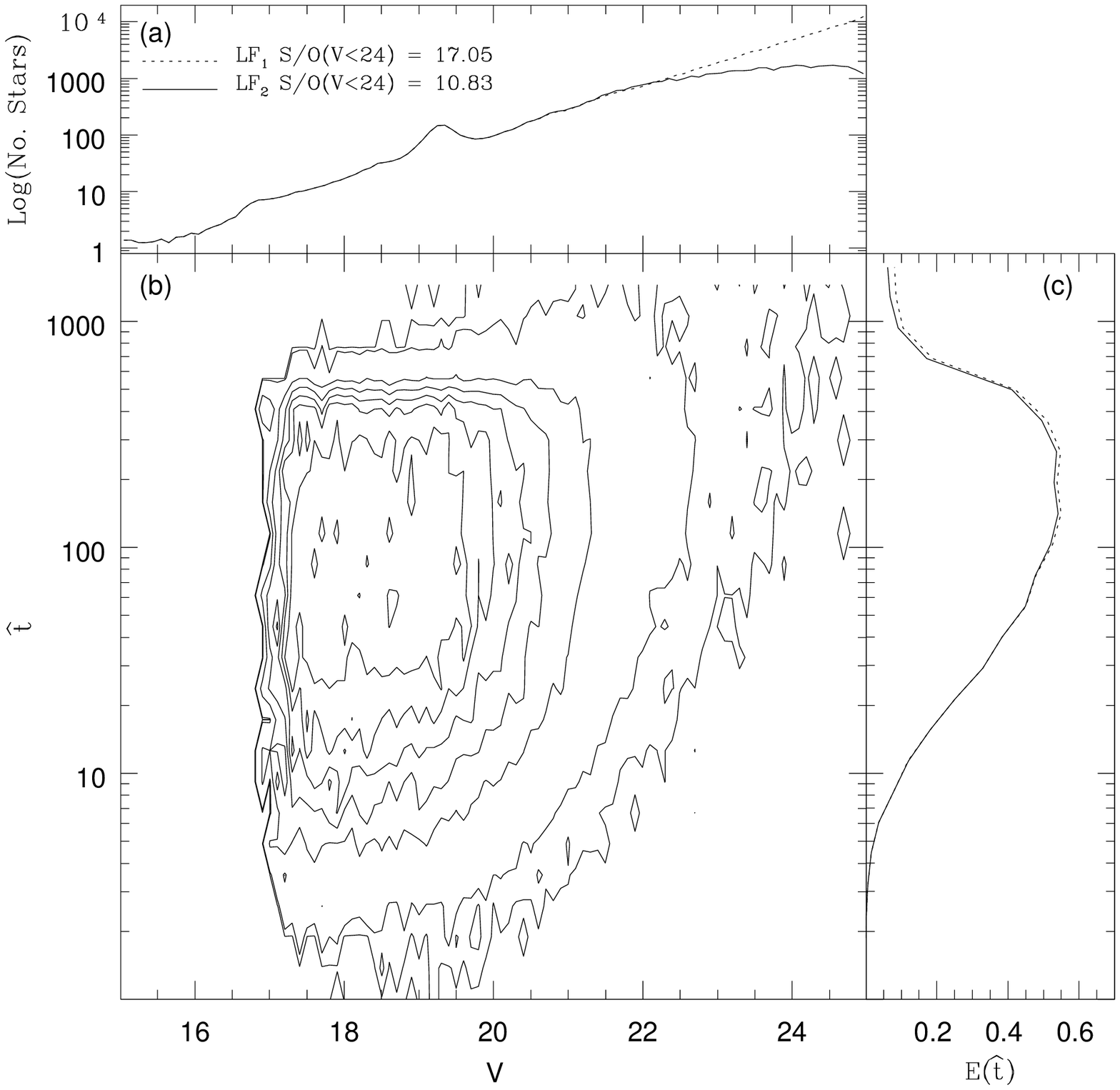}
\figcaption[fig_eff_per_bin_cutB.ps]{
Panel (a) displays luminosity functions $LF_{1}$
(\textit{dotted line}) and $LF_{2}$ (\textit{solid line}), which
are used to integrate $\epsilon(\that,V)$ over $V$.  The 
function $\epsilon(\that,V)$ is shown in
panel (b) as a contour plot, with the contours corresponding to
efficiencies of 
0.001, 0.01, 0.1, 0.2, 0.3, 0.4, and 0.5.  The function $\epsilon(\that,V)$
is described more fully in the text.  The result of integrating 
$\epsilon(\that,V)$ over the two LFs and scaling by the
appropriate normalization to obtain $\Ethat$ is shown in
panel (c).  The line coding corresponds to that in panel (a).  Note 
that $\Ethat$ is fairly insensitive to the shape of the luminosity
function for dim stars because, (1) the function $\epsilon(\that,V)$ 
is small here and (2) although the faint stars are weighted less in
$LF_{2}$ they also contribute less to the normalization $S/O(V<24)$.
\label{fig_eff_per_bin}}

\clearpage

\plotone{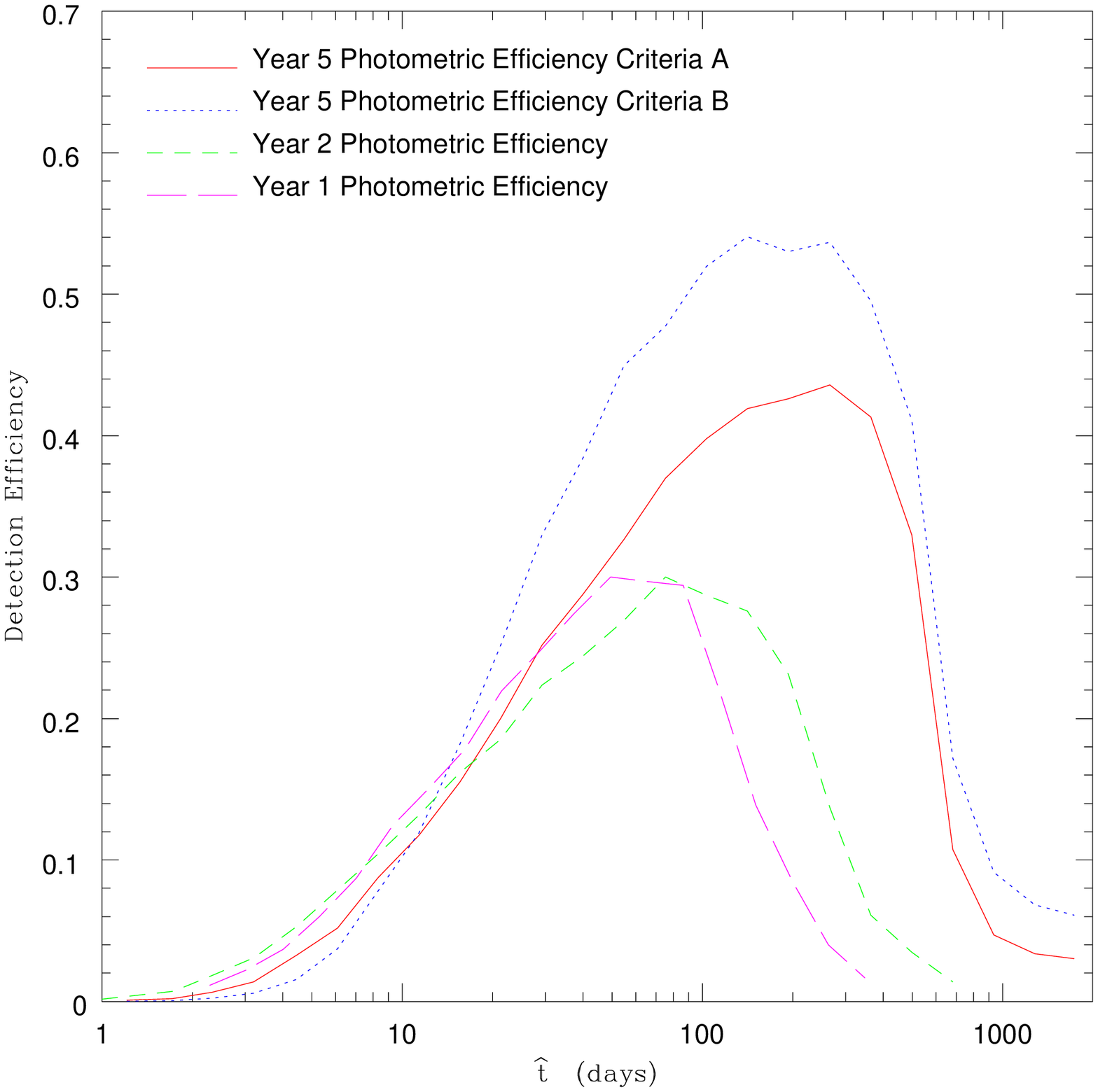}
\figcaption[fig_phot_eff.ps]{
Microlensing detection efficiency (normalized to $\umin < 1$)
for the 5.7-year MACHO data, as a function of event timescale $\that$.
The \textit{solid line} shows the `photometric' efficiency computed for 
criteria set A, and the \textit{dotted line} for criteria set B as described
in \S~\protect\ref{subsec-phoeff}.  For comparison the corresponding curves
from \yrone\ and \yrtwo\ are also shown.
\label{fig_phot_eff}}

\plotone{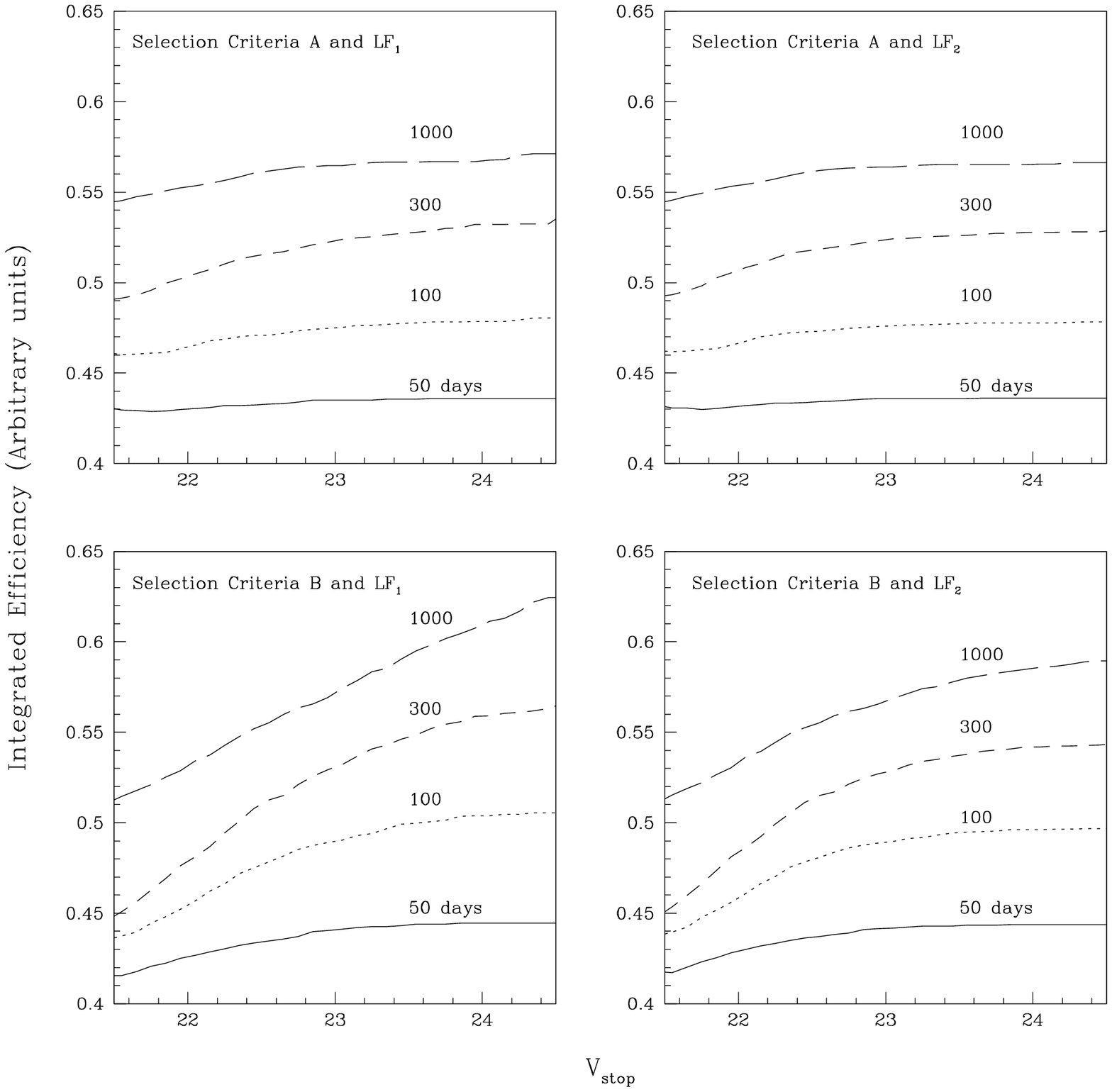}
\figcaption[fig_convergence4.ps]{
Convergence of $\Ethat$ with $V_{\rm{stop}}$ is shown for 
several different durations, $\that$, and for the four 
combinations of selection criteria sets A and B and luminosity
functions $LF_{1}$ and $LF_{2}$.
\label{fig_convergence}}

\plotone{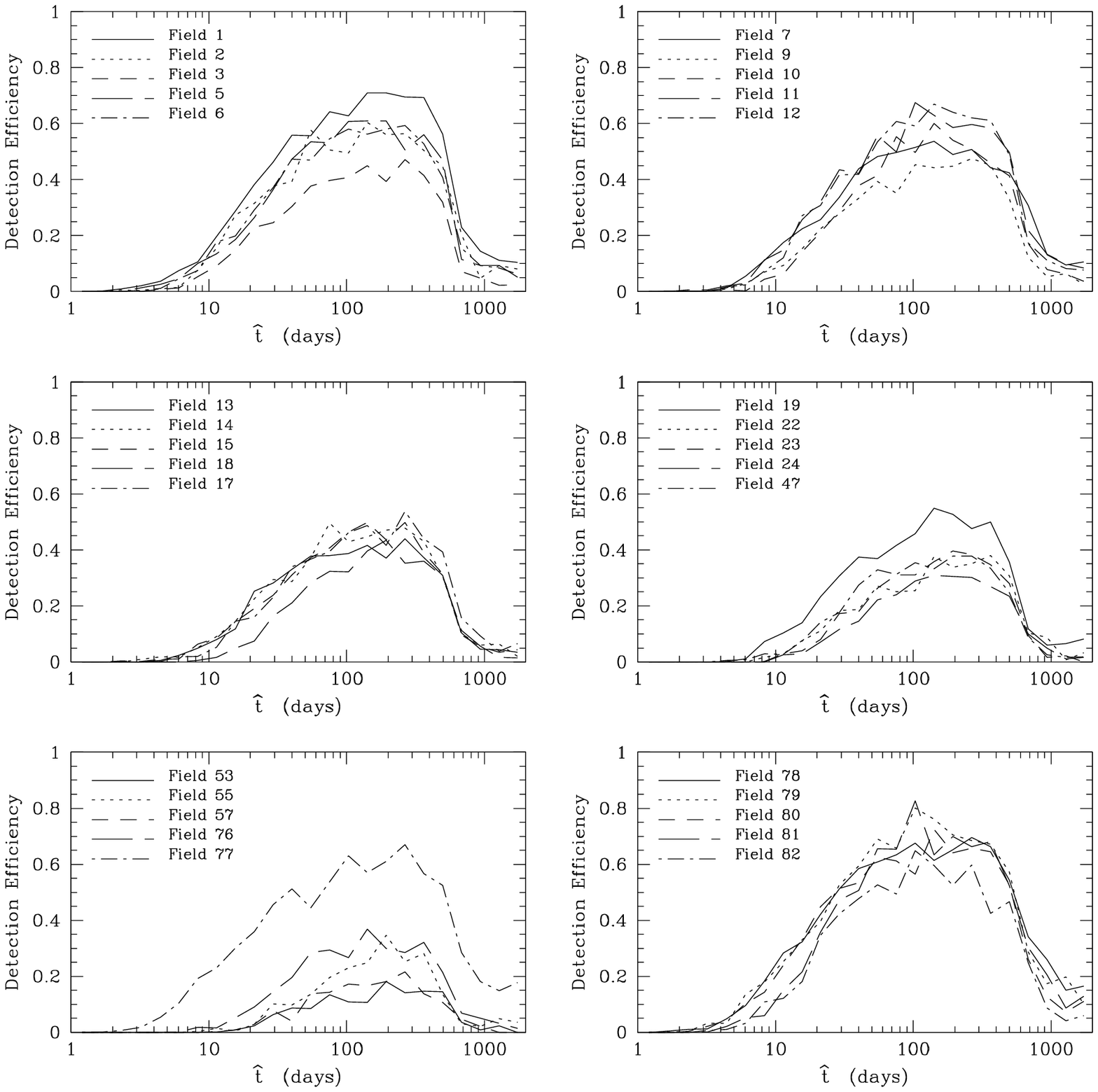}
\figcaption[fig_photeff_vs_field_cutB.ps]{
The photometric efficiency for each field is displayed for the 30
fields analyzed in \yrfive.  The different sampling rates, stellar 
densities, and crowding all contribute to the
differences seen between fields.
\label{fig_photeff_vs_field}}

\plotone{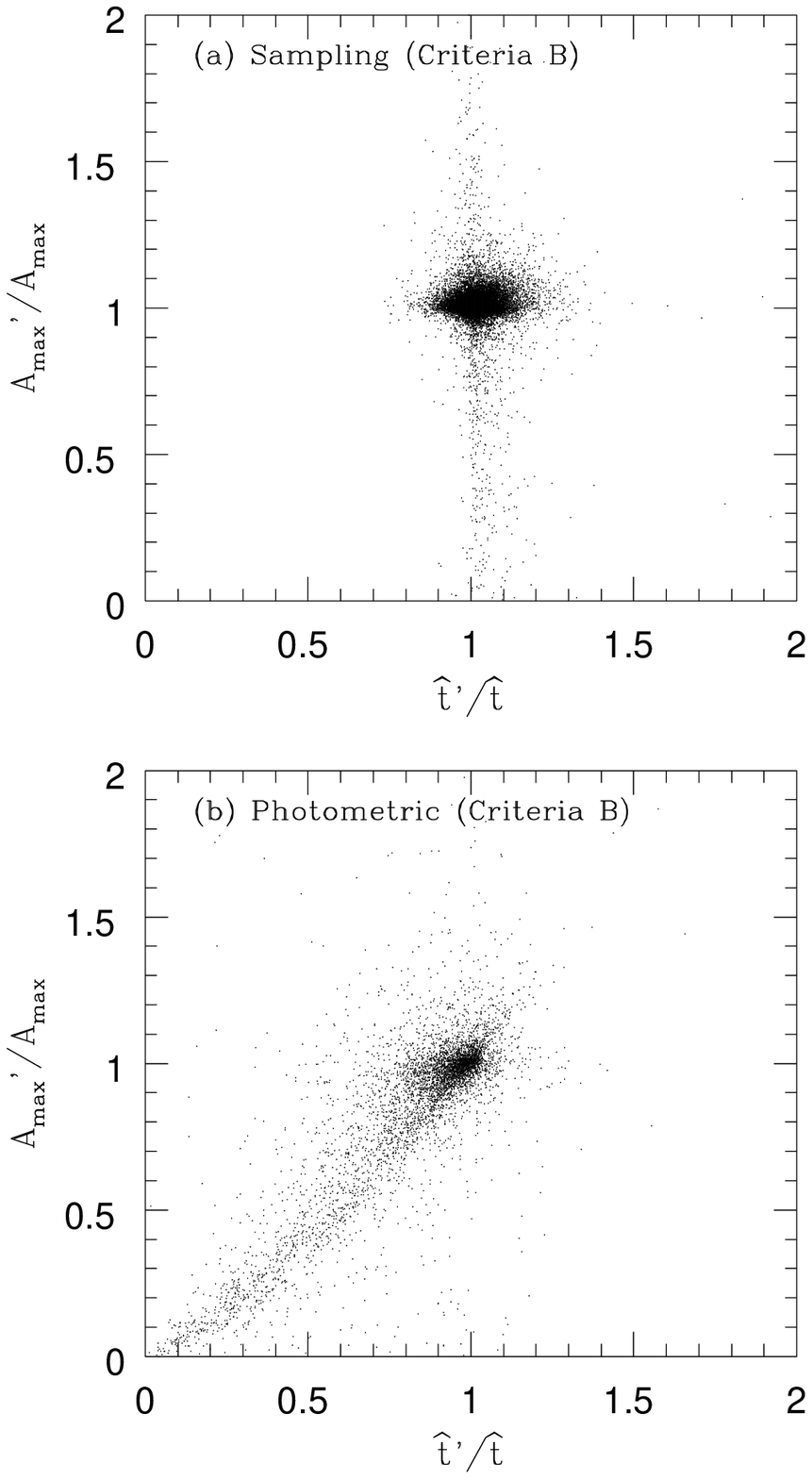}
\figcaption[fig_bias.ps]{
Parameter biases in the fitting code are induced by blending as
illustrated here in a scatter plot of $\rAmax/\Amax$ versus
$\rthat/\that$ for a sample of Monte Carlo events.  Panel (a)
is a sample created with the sampling efficiency methods 
and panel (b) is a sample created using the photometric efficiency
methods (both use selection criteria set B).
\label{fig_bias}}

\plotone{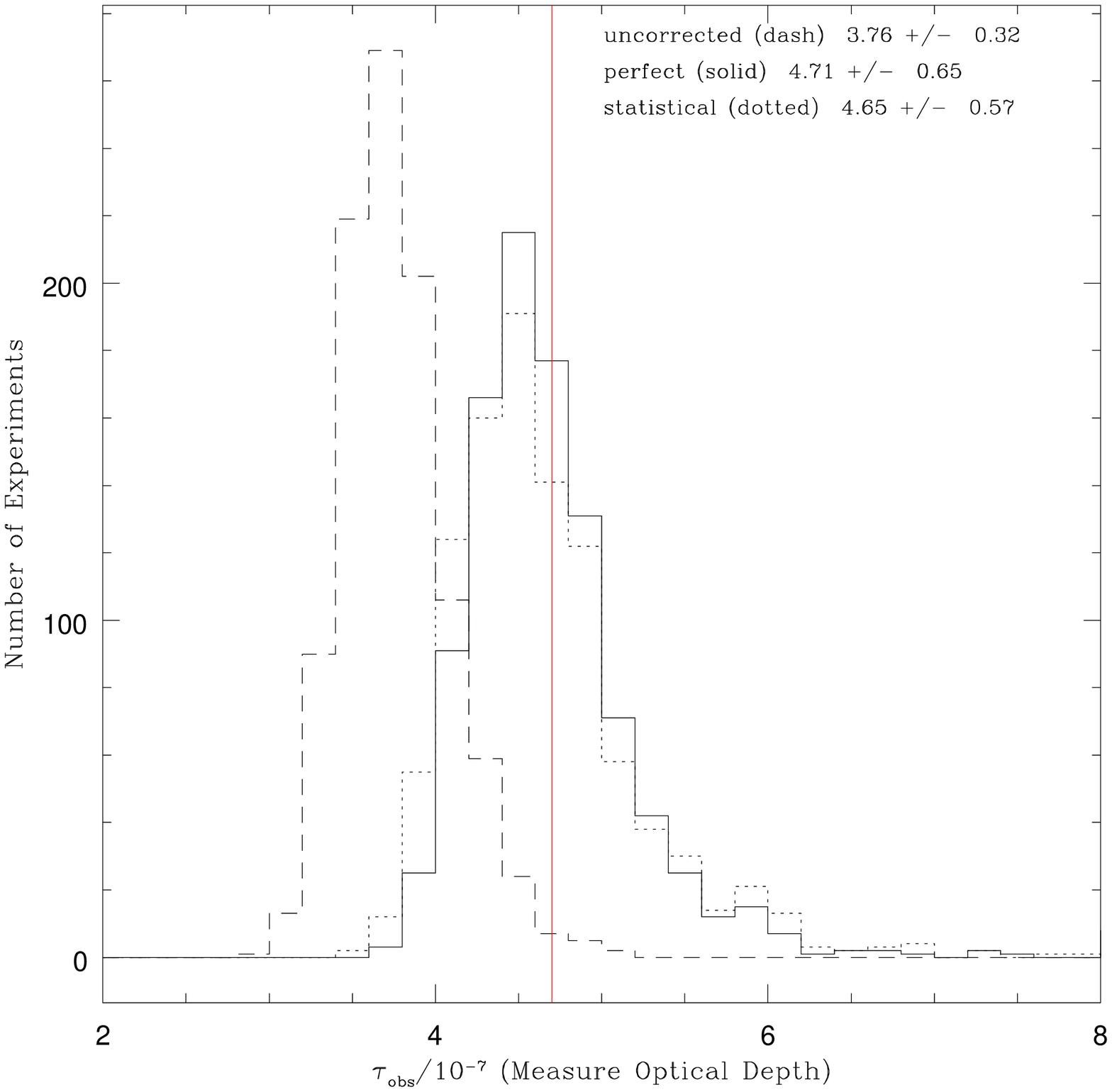}
\figcaption[fig_monte_tau_cutB.ps]{
The results of a Monte Carlo $\that$ bias test described in the 
text is shown. 
Here we have used $LF_{2}$ and 
criteria set B along with a value of the bias correction 
of $\overline{\alpha} = 0.785$ (Table~\ref{tab-bias}) computed
using Equation~\ref{equation18}.  The number of expected events
per trial was $N_{exp} = 64$.
The \textit{dashed histogram} is the distribution in $\tau_{meas}$ using the 
uncorrected $\rthat$ to compute the optical depth, and as expected 
is heavily biased toward smaller optical depths.  The \textit{solid 
histogram} is the distribution using the true values of $\that$ 
and is exact at recovering the optical depth on average.
The \textit{dotted histogram} is the distribution in $\tau_{meas}$ 
using the statistically corrected values of 
$\that_{st}\,\rightarrow\,\rthat/\overline{\alpha}$.  
Using $\overline{\alpha}$ has corrected for the $\that$ bias
without significantly increasing the spread in optical depths.
\label{fig_monte_tau}}

\twocolumn
\newpage

\begin{deluxetable}{rccccccccc}  
\tablecaption{ Normalizations \label{tab-norms} }
\tablewidth{0pt}
\tablehead{
\colhead{Field} &
\colhead{air} &
\colhead{seeing} &
\colhead{sky} &
\colhead{offset} &
\colhead{Density} &
\colhead{$LF_{1}$} &
\colhead{$LF_{1}$} &
\colhead{$LF_{2}$} &
\colhead{$LF_{2}$}
\nl
\colhead{} &
\colhead{} &
\colhead{(")} &
\colhead{($e^{-}$)} &
\colhead{(V mag)} &
\colhead{($Objs/\Box'$)} &
\colhead{$S/O(22)$} &
\colhead{$S/O(24)$} &
\colhead{$S/O(22)$} &
\colhead{$S/O(24)$}
}  
\startdata
  1 & 2.01 & 1.68 & 2740.0 &  0.10 & 241.3 & $3.6_{-0.7}^{+0.6}$ & $22.6_{-4.6}^{+4.0}$ & $3.7_{-0.8}^{+0.7}$ & $14.3_{-2.9}^{+2.5}$ \nl
  2 & 1.56 & 1.63 & 3423.0 &  0.08 & 184.7 & $2.7_{-0.5}^{+0.5}$ & $17.1_{-3.5}^{+3.1}$ & $2.8_{-0.6}^{+0.5}$ & $10.9_{-2.2}^{+1.9}$ \nl
  3 & 1.56 & 1.53 & 2950.0 & -0.02 & 173.4 & $1.8_{-0.3}^{+0.3}$ & $11.1_{-2.1}^{+1.8}$ & $1.8_{-0.3}^{+0.3}$ & $7.1_{-1.3}^{+1.1}$ \nl
  5 & 1.21 & 1.78 & 3285.0 &  0.12 & 237.9 & $2.7_{-0.6}^{+0.5}$ & $17.0_{-3.5}^{+3.2}$ & $2.8_{-0.6}^{+0.5}$ & $10.8_{-2.2}^{+2.0}$ \nl
  6 & 1.53 & 1.52 & 3176.0 &  0.08 & 226.4 & $2.5_{-0.5}^{+0.4}$ & $15.9_{-3.2}^{+2.8}$ & $2.6_{-0.5}^{+0.5}$ & $10.1_{-2.0}^{+1.8}$ \nl
  7 & 1.46 & 1.60 & 3122.0 &  0.25 & 233.7 & $2.8_{-0.6}^{+0.5}$ & $17.8_{-3.8}^{+3.3}$ & $2.9_{-0.6}^{+0.5}$ & $11.3_{-2.4}^{+2.1}$ \nl
  9 & 1.33 & 1.49 & 1771.0 &  0.12 & 209.7 & $1.9_{-0.4}^{+0.3}$ & $11.9_{-2.5}^{+2.1}$ & $2.0_{-0.4}^{+0.4}$ & $7.6_{-1.6}^{+1.4}$ \nl
 10 & 1.31 & 1.66 & 2450.0 &  0.12 & 182.3 & $2.4_{-0.5}^{+0.4}$ & $14.9_{-3.1}^{+2.7}$ & $2.4_{-0.5}^{+0.4}$ & $9.5_{-2.0}^{+1.7}$ \nl
 11 & 1.46 & 1.59 & 3153.0 &  0.10 & 221.9 & $2.8_{-0.6}^{+0.5}$ & $17.4_{-3.5}^{+3.0}$ & $2.8_{-0.6}^{+0.5}$ & $11.0_{-2.2}^{+1.9}$ \nl
 12 & 1.57 & 1.58 & 3109.0 & -0.14 & 202.9 & $2.7_{-0.5}^{+0.5}$ & $16.8_{-3.4}^{+3.0}$ & $2.8_{-0.6}^{+0.5}$ & $10.7_{-2.2}^{+1.9}$ \nl
 13 & 1.32 & 1.55 & 2290.0 &  0.10 & 179.6 & $1.7_{-0.4}^{+0.3}$ & $10.7_{-2.3}^{+1.9}$ & $1.8_{-0.4}^{+0.3}$ & $6.8_{-1.4}^{+1.2}$ \nl
 14 & 1.32 & 1.62 & 1636.0 &  0.12 & 211.1 & $1.8_{-0.4}^{+0.3}$ & $11.3_{-2.2}^{+1.9}$ & $1.9_{-0.4}^{+0.3}$ & $7.2_{-1.4}^{+1.2}$ \nl
 15 & 1.32 & 1.80 & 1543.0 & -0.08 & 182.9 & $1.9_{-0.4}^{+0.4}$ & $12.2_{-2.5}^{+2.3}$ & $2.0_{-0.4}^{+0.4}$ & $7.8_{-1.6}^{+1.5}$ \nl
 17 & 1.36 & 1.55 & 2250.0 &  0.15 & 176.9 & $2.3_{-0.4}^{+0.4}$ & $14.2_{-2.7}^{+2.3}$ & $2.3_{-0.4}^{+0.4}$ & $9.0_{-1.7}^{+1.5}$ \nl
 18 & 1.34 & 1.51 & 2310.0 &  0.15 & 214.0 & $2.1_{-0.4}^{+0.4}$ & $13.3_{-2.7}^{+2.3}$ & $2.2_{-0.4}^{+0.4}$ & $8.5_{-1.7}^{+1.5}$ \nl
 19 & 2.50 & 1.80 & 2951.0 &  0.17 & 169.6 & $2.5_{-0.5}^{+0.5}$ & $15.8_{-3.2}^{+2.9}$ & $2.6_{-0.5}^{+0.5}$ & $10.1_{-2.0}^{+1.8}$ \nl
 22 & 1.38 & 1.71 & 2250.0 &  0.17 & 141.1 & $1.7_{-0.3}^{+0.3}$ & $10.7_{-2.2}^{+1.9}$ & $1.8_{-0.4}^{+0.3}$ & $6.8_{-1.4}^{+1.2}$ \nl
 23 & 1.38 & 1.65 & 2280.0 &  0.19 & 158.4 & $2.1_{-0.4}^{+0.4}$ & $13.1_{-2.5}^{+2.2}$ & $2.2_{-0.4}^{+0.4}$ & $8.3_{-1.6}^{+1.4}$ \nl
 24 & 1.35 & 2.00 & 2090.0 &  0.25 & 148.3 & $1.7_{-0.3}^{+0.3}$ & $10.5_{-2.1}^{+1.9}$ & $1.7_{-0.4}^{+0.3}$ & $6.7_{-1.4}^{+1.2}$ \nl
 47 & 1.60 & 1.57 & 2440.0 &  0.15 & 141.7 & $1.5_{-0.3}^{+0.3}$ & $9.7_{-2.0}^{+1.7}$ & $1.6_{-0.3}^{+0.3}$ & $6.2_{-1.3}^{+1.1}$ \nl
 53 & 2.14 & 1.79 & 2840.0 &  0.40 & 100.0 & $1.6_{-0.3}^{+0.3}$ & $10.0_{-2.0}^{+1.9}$ & $1.6_{-0.3}^{+0.3}$ & $6.4_{-1.3}^{+1.2}$ \nl
 55 & 2.01 & 1.83 & 2590.0 &  0.23 &  93.0 & $1.9_{-0.4}^{+0.3}$ & $11.7_{-2.4}^{+2.2}$ & $1.9_{-0.4}^{+0.4}$ & $7.4_{-1.5}^{+1.4}$ \nl
 57 & 1.84 & 1.89 & 2510.0 &  0.29 &  98.9 & $1.6_{-0.3}^{+0.3}$ & $9.8_{-2.0}^{+1.8}$ & $1.6_{-0.3}^{+0.3}$ & $6.2_{-1.2}^{+1.1}$ \nl
 76 & 1.25 & 1.60 & 2470.0 &  0.15 & 167.0 & $1.8_{-0.4}^{+0.3}$ & $11.1_{-2.2}^{+2.0}$ & $1.8_{-0.4}^{+0.3}$ & $7.1_{-1.4}^{+1.2}$ \nl
 77 & 1.39 & 1.48 & 2314.0 &  0.06 & 316.3 & $4.4_{-0.9}^{+0.8}$ & $27.9_{-5.8}^{+5.1}$ & $4.6_{-1.0}^{+0.8}$ & $17.8_{-3.7}^{+3.2}$ \nl
 78 & 1.41 & 1.58 & 3545.0 &  0.12 & 290.2 & $4.6_{-1.0}^{+0.8}$ & $28.8_{-6.0}^{+5.3}$ & $4.7_{-1.0}^{+0.9}$ & $18.3_{-3.8}^{+3.3}$ \nl
 79 & 1.43 & 1.65 & 3465.0 &  0.08 & 255.5 & $4.3_{-0.9}^{+0.8}$ & $26.7_{-5.5}^{+4.7}$ & $4.4_{-0.9}^{+0.8}$ & $17.0_{-3.5}^{+3.0}$ \nl
 80 & 1.43 & 1.56 & 3574.0 &  0.00 & 237.9 & $3.4_{-0.7}^{+0.6}$ & $21.6_{-4.4}^{+3.9}$ & $3.5_{-0.7}^{+0.6}$ & $13.7_{-2.8}^{+2.5}$ \nl
 81 & 1.52 & 1.70 & 3657.0 & -0.04 & 210.6 & $4.0_{-0.8}^{+0.7}$ & $25.2_{-5.1}^{+4.5}$ & $4.1_{-0.8}^{+0.7}$ & $16.0_{-3.2}^{+2.8}$ \nl
 82 & 1.52 & 1.45 & 3580.0 & -0.04 & 170.4 & $3.0_{-0.6}^{+0.5}$ & $18.9_{-3.6}^{+3.1}$ & $3.1_{-0.6}^{+0.5}$ & $12.0_{-2.3}^{+2.0}$ \nl

 \enddata
\tablenotetext{} {
The table lists the normalizations (columns 7-10) for each of the 30 fields,
along with conservative estimated errors.  The columns 2-4 list the airmass,
seeing (in arcsecs), and sky level (in electrons) for the MACHO template
images.  Column 5 is the offset for each field used in \S~\ref{sec-lfnorm}.
Column 6 is the average density of each field in $Objects/\Box'$.}
\end{deluxetable}
\newpage

\begin{deluxetable}{rrrccccc}  
\tablecaption{ PDRB Chunks \label{tab-chunks} }
\tablewidth{0pt}
\tablehead{
\colhead{ID} & \colhead{Field} & \colhead{Chunk} &
\colhead{Template} & \colhead{$Objs/\Box'$} & \colhead{$N_{cond}$} &
\colhead{$N_{PRF}/grid$} & \colhead{$N_{PRF}$}
}  

\startdata
 1 &  2 &  9 & 5672 & 143.9 & 67 &  64 & 3226 \nl
 2 &  7 & 40 & 5668 & 196.4 & 66 &  81 & 4209 \nl
 3 & 19 & 22 & 4562 & 142.1 & 72 &  64 & 3219 \nl
 4 & 22 & 38 & 1696 &  93.7 & 64 &  64 & 3419 \nl
 5 & 55 & 54 & 5311 &  98.1 & 69 &  64 & 3294 \nl
 6 & 77 & 27 & 4203 & 257.4 & 67 & 144 & 6884 \nl
 7 & 77 & 50 & 2962 & 308.9 & 66 & 156 & 7622 \nl
 8 & 78 &  6 & 5663 & 318.7 & 69 & 156 & 8381 \nl
 9 & 78 & 47 & 5663 & 269.5 & 77 & 144 & 8074 \nl
10 & 81 & 18 & 5673 & 202.9 & 72 & 156 & 6653 \nl

\enddata
\tablenotetext{} {
The ten $5'\!\times\!5'$ chunks used
to create the PRDB span the observed range of density
($Objs/\Box'$) and are widely spaced over the face of the LMC.
For reference we give the template observation number,
the number of observing conditions ($N_{cond}$) used for each
chunk, the number of artificial stars per grid ($N_{PRF}/grid$)
added to each chunk (60 grids per chunk) and the number of 
useful PRFs generated ($N_{PRF}$) from each chunk.}
\end{deluxetable}

\newpage

\begin{deluxetable}{cccccc}  
\tablecaption{ Efficiency Table \label{tab-eff} }
\tablewidth{0pt}
\tablehead{
\colhead{Bin} &
\colhead{$\that_{min}$} &
\colhead{$\that_{max}$} &
\colhead{$\that$} &
\colhead{Criteria set A} &
\colhead{Criteria set B}
\nl
\colhead{} &
\colhead{} &
\colhead{} &
\colhead{} &
\colhead{$\Ethat$} &
\colhead{$\Ethat$}
}  
\startdata
%
  1  &    1.00 &    1.46 &    1.20 & 8.7e-04  & 1.9e-04 \nl
  2  &    1.46 &    2.00 &    1.70 & 2.1e-03  & 5.7e-04 \nl
  3  &    2.00 &    2.70 &    2.32 & 6.5e-03  & 2.2e-03 \nl
  4  &    2.70 &    3.80 &    3.20 & 1.3e-02  & 5.7e-03 \nl
  5  &    3.80 &    5.20 &    4.44 & 3.2e-02  & 1.5e-02 \nl
  6  &    5.20 &    7.10 &    6.07 & 5.2e-02  & 3.7e-02 \nl
  7  &    7.10 &    9.70 &    8.29 & 8.7e-02  & 7.8e-02 \nl
  8  &    9.70 &   13.30 &   11.35 & 1.1e-01  & 1.1e-01 \nl
  9  &   13.30 &   18.20 &   15.55 & 1.5e-01  & 1.8e-01 \nl
 10  &   18.20 &   24.90 &   21.28 & 1.9e-01  & 2.5e-01 \nl
 11  &   24.90 &   34.20 &   29.18 & 2.5e-01  & 3.2e-01 \nl
 12  &   34.20 &   46.80 &   40.00 & 2.8e-01  & 3.8e-01 \nl
 13  &   46.80 &   64.20 &   54.81 & 3.2e-01  & 4.4e-01 \nl
 14  &   64.20 &   88.00 &   75.16 & 3.6e-01  & 4.7e-01 \nl
 15  &   88.00 &  121.00 &  103.18 & 3.9e-01  & 5.1e-01 \nl
 16  &  121.00 &  165.00 &  141.29 & 4.1e-01  & 5.4e-01 \nl
 17  &  165.00 &  227.00 &  193.53 & 4.2e-01  & 5.3e-01 \nl
 18  &  227.00 &  311.00 &  265.70 & 4.3e-01  & 5.3e-01 \nl
 19  &  311.00 &  426.00 &  363.98 & 4.1e-01  & 4.9e-01 \nl
 20  &  426.00 &  584.00 &  498.78 & 3.2e-01  & 4.1e-01 \nl
 21  &  584.00 &  800.00 &  683.52 & 1.0e-01  & 1.7e-01 \nl
 22  &  800.00 & 1096.00 &  936.37 & 4.7e-02  & 9.1e-02 \nl
 23  & 1096.00 & 1502.00 & 1283.04 & 3.3e-02  & 6.8e-02 \nl
 24  & 1502.00 & 2000.00 & 1733.14 & 3.0e-02  & 6.1e-02 \nl
\enddata
\tablenotetext{}{
Efficiencies presented in \yrfive\ and also shown in
Figure~\ref{fig_phot_eff} for
selection criteria set A and B.  There are 24 bins in duration with
$\that_{min}$ and $\that_{max}$ delineating each bin.  The quantity
$\that$ marks the logarithmic center of each bin.}
\end{deluxetable}

\newpage

\begin{deluxetable}{clrl}  
\tablecaption{ Error Budget \label{tab-errorbudget} }
\tablewidth{0pt}
\tablehead{
\colhead{Number} &
\colhead{Sign} &
\colhead{Size} &
\colhead{Note}
}  
\startdata
1  &  ?  &  $\sim$20\%    & LF normalization/exposure           \nl
2  &  ?  &  $\sim$1-3\%   & LF for faint stars $V\simgt22.5$    \nl
3  &  ?  &  $\simlt$1\%   & Finite number of Monte Carlo events \nl
4  &  +  &  $\sim$6-8\%   & Overcompleteness bias               \nl
5  &  -  &  $\sim$3\%     & 1st neighbor PRF bias               \nl
6  &  +  &  $\sim$1\%     & LF in outer `disk' fields           \nl
7  &  -  &  $\simlt$1\%   & Convergence for $\that > 100$ days  \nl
8  &  ?  &       ?\%      & Binary stars in LF                  \nl
9  &  ?  &       ?\%      & No exotic lensing events            \nl
\enddata
\tablenotetext{} {
This table details our error budget for the efficiency.
Column 2 gives the probable direction, if known, in which the error
might propagate to the efficiency, \eg, \#4 could bias the efficiency to
higher values, causing the optical depth to be slightly underestimated.
Two possible corrections that are not estimated here are \#8 and \#9
(see the text).}
\end{deluxetable}

\newpage

\begin{deluxetable}{ccccc}  
\tablecaption{ $\that$ Bias Corrections \label{tab-bias} }
\tablewidth{0pt}
\tablehead{
\colhead{Selection} &
\colhead{LF } &
\colhead{ } &
\colhead{Median $\overline{\alpha}$} &
\colhead{ }
\nl
\colhead{Criteria} &
\colhead{ } &
\colhead{0.1 $M_{\odot}$} &
\colhead{0.5 $M_{\odot}$} &
\colhead{1.0 $M_{\odot}$}
}  
\startdata
 A     &     1   &    0.864  &  0.826  &  0.812  \nl
 A     &     2   &    0.867  &  0.831  &  0.818  \nl
 B     &     1   &    0.826  &  0.777  &  0.759  \nl
 B     &     2   &    0.828  &  0.785  &  0.770  \nl
%
%
\enddata
\tablenotetext{} {This table lists $\that$ bias corrections
which may be used to statistically correct the unblended
fit $\that$'s.  This statistical correction is a function of
the selection criteria set used, but also sensitive to the average
duration of the observed microlensing events, here delineated
using a standard Galactic halo model (model S) with delta-function
masses of 
0.1, 0.5, and 1.0 $M_{\odot}$, corresponding to mean durations
of 41, 92, and 130 days, respectively.  We recommend using
$LF_{2}$ and 0.5 $M_{\odot}$.}
\end{deluxetable}


\clearpage

\begin{thebibliography}{}

\bibitem[Afonso \etal\ 1998]{smc-binary} 
  Alfonso, C. \etal 1998, \aap, 337, L17

\bibitem[Afonso \etal\ 1999]{erossmc2} 
  Alfonso, C. \etal 1999, \aap, 344, L63

\bibitem[Albrow \etal\ 1998a]{albrow-planet} 
  Albrow, M.D. \etal 1998a, \apj, 509, 687

\bibitem[Albrow \etal\ 1998b]{planetsmc2} 
  Albrow, M.D. \etal 1998b, \apj, 512, 672

\bibitem[Alcock \etal\ 1995]{explore}
  Alcock, C. \etal 1995, \apj, 449, 28

\bibitem[Alcock \etal\ 1996]{lmc1}
  Alcock, C. \etal 1996, \apj, 461, 84

\bibitem[Alcock \etal\ 1997a]{lmc2}
  Alcock, C. \etal 1997a, \apj, 486, 697

\bibitem[Alcock \etal\ 1997b]{smc1}
  Alcock, C. \etal 1997b, \apjl, 491, L11 

\bibitem[Alcock \etal\ 1997c]{95-30}
  Alcock, C. \etal 1997c, \apj, 491, 436

\bibitem[Alcock \etal\ 1999a]{smc2}
  Alcock, C. \etal 1999a, \apj, 518, 44

\bibitem[Alcock \etal\ 1999b]{calib}
  Alcock, C. \etal  1999b, PASP, 111, 1539

\bibitem[Alcock \etal\ 1999c]{bulge-binary}
  Alcock, C. \etal 1999c, \apj, submitted \\
  (astro-ph/9907369)

\bibitem[Alcock \etal\ 1999d]{9mcmd}
  Alcock, C. \etal 1999d, \aj, submitted \\
  (astro-ph/0001435)

\bibitem[Alcock \etal\ 1999e]{lmc5}
  Alcock, C. \etal 1999e, \apj, submitted \\
  (astro-ph/0001272)

\bibitem[Ansari et al. 1997]{AGAPE}  
  Ansari, R. 1997, A\&A, 324, 843

\bibitem[Crotts et al. 1999]{MEGA99}  
  Crotts A.P.S., Uglesich, R., Gyuk, G., Tomaney, A.B., 1999,
  Astronomical Society of the Pacific Conference Series, 1999, 182, 409

\bibitem[Di Stefano \& Perna 1997]{distefano1} 
  Di Stefano, R. and Perna, R., 1997 \apj, 488, 55

\bibitem[Di Stefano 1999]{distefano2}   
  Di Stefano, R., 1999, \apjl, submitted \\
  (astrop-ph/9901035)

\bibitem[Geha \etal\ 1998]{geha}  
  Geha, M. \etal 1998, \apj, 115, 1045

\bibitem[Griest 1991] {griest91}
  Griest,~K., 1991, \apj, 366, 412

\bibitem[Griest \& Hu 1991]{griest-hu} 
  Griest, K. and Hu, W., 1992, \apj, 397, 362

\bibitem[Groth 1986]{groth}
  Groth, E.J. 1986, \apj, 91, 1244

\bibitem[Han 1997]{hanhubble} 
  Han, C. 1997, \apj, 490, 51

\bibitem[Han 1998]{han} 
  Han, C. 1998, \apj, 500, 569

\bibitem[Kroupa 1995]{kroupa} 
  Kroupa, P., 1995, \apj, 453, 358

\bibitem[Lasserre \etal\ 2000] {eros-2}   
 Lasserre, T. \etal\, 2000, A\&A Letters, in press \\
 (astro-ph/0002253)

\bibitem[Mao \& Paczy{\'n}ski 1991] {mao-pac} 
  Mao,~S. and Paczy{\'n}ski,~B., 1991, \apjl, 374, L37

\bibitem[Marshall \etal\ 1994]{marshall}  
  Marshall, S.L. \etal 1994, in IAU Symp. 161, 67

\bibitem[Olsen \etal\ 1998]{olsen1} 
  Olsen, K.A.G., Hodge, P.W., Mateo, M., Olszewski, E.W.,
  Schommer, R.A., Suntzeff, N.B., Walker, A.R., 
  \etal 1998, \mnras, 300, 665

\bibitem[Olsen 1999]{olsen2}  
  Olsen, K.A.G. 1999, \apj, 117, 2244

\bibitem[Palanque-Delabrouille \etal\ 1998]{erossmc1} 
  Palanque-Delabrouille, N. \etal 1999, \aap, 332, 1

\bibitem[Reid \&  Gizis 1997]{reid}   
  Reid,N.I. and Gizis, J.E., 1997, \aj, 113, 2246

\bibitem[Rhie \etal 1999]{rhie-mps}   
  Rhie, S.H., Becker, A.C., Bennett, D.P., Fragile, P.C.,
  Johnson, B.R., King, L.J., Peterson, B.A., Quinn, J.,
  \etal 1999, \apj, 522, 1037 

\bibitem[Pratt 1997]{pratt}  
  Pratt, M. 1997, PhD Thesis, University of California at Santa Barbara
  \& University of Washington

\bibitem[Roulet \& Mollerach 1997]{roulet}  
  Roulet, E. and Mollerach, S. 1997, Physics Reports, 279, 68

\bibitem[Schechter, Mateo, \& Saha 1993]{dophot} 
  Schechter, P.L., Mateo, M., and Saha, A. 1993, PASP, 105, 1342

\bibitem[Stubbs \etal\ 1993]{stubbs}  
  Stubbs, C.W. \etal 1993, Proc. SPIE, 1900, 192

\bibitem[Stetson 1992]{stetson}  
  Stetson, P.B. 1992, ASP Conf. Ser. 25, 297

\bibitem[Udalski \etal\ 1997]{udalski-erosII}  
  Udalski, A. \etal 1997, Acta Astronomica, 47, 319  

\bibitem[Wozniak 1997]{wozniak}  
  Wozniak, P. and Paczy{\'n}ski, B. 1997, \apj, 487, 55 

\end{thebibliography}
\end{document}